\documentstyle[preprint,aps,epsf]{revtex}

\def\P{{\sf \bf P}}
\def\R{{\sf \bf R}}
\def\W{{\sf \bf W}}
\def\C{{\sf \bf C}}
\def\PW{{\sf \bf PW}}
\def\PC{{\sf \bf PC}}
\def\RW{{\sf \bf RW}}
\def\RC{{\sf \bf RC}}
\def\PCtr{{\sf \bf  PC$_{\rm tree}$}}
\def\PCmf{{\sf \bf  PC$_{\rm MF}$}}

\def\RCtr{{\sf \bf  RC$_{\rm tree}$}}
\def\RCmf{{\sf \bf  RC$_{\rm MF}$}}
\def\RCpmf{{\sf \bf RC$_{\rm pMF}$}}

\begin{document}

\draft
\tightenlines

\title{
%
% Preprint Number
%
\begin{flushright}
\normalsize
UTCCP-P-61          \\
\end{flushright}
%
% Title
%
Comparative Study of full QCD Hadron Spectrum and 
Static Quark Potential with Improved
Actions}  

\author{CP-PACS Collaboration  \\[1mm] 
        S.~Aoki$^{\rm a}$, 
        G.~Boyd$^{\rm b}$, 
        R.~Burkhalter$^{\rm a,b}$,
        S.~Hashimoto$^{\rm c}$, 
        N.~Ishizuka$^{\rm a,b}$,
        Y.~Iwasaki$^{\rm a,b}$, 
        K.~Kanaya$^{\rm a,b}$, 
        T.~Kaneko$^{\rm b}$, 
        Y.~Kuramashi$^{\rm d}$
\thanks{On leave from Institute of Particle and Nuclear Studies,
High Energy Accelerator Research Organization(KEK),
Tsukuba, Ibaraki 305-0801, Japan},
        M.~Okawa$^{\rm e}$,
        A.~Ukawa$^{\rm a,b}$, 
        T.~Yoshi\'{e}$^{\rm a,b}$ }

\address{$ ^a$Institute of Physics, 
         University of Tsukuba,\\ 
         Tsukuba, Ibaraki 305-8571, Japan\\
         $ ^b$Center for Computational Physics, 
         University of Tsukuba,\\ 
         Tsukuba, Ibaraki 305-8577, Japan\\
         $ ^c$Computing Research Center, 
         High Energy Accelerator Research Organization(KEK), \\
         Tsukuba, Ibaraki 305-0801, Japan\\
         $ ^d$Department of Physics, Washington University, 
         St. Louis, Missouri 63130, USA\\
         $ ^e$Institute of Particle and Nuclear Studies, 
         High Energy Accelerator Research Organization(KEK),\\ 
         Tsukuba, Ibaraki 305-0801, Japan}

\date{February 1999}

\maketitle

\newpage

\begin{abstract}

We investigate effects of action improvement on the light hadron
spectrum and the static quark potential in two-flavor QCD for $a^{-1} \approx
1$~GeV and $m_{\rm PS}/m_{\rm V} = 0.7$--0.9.  
We compare a renormalization group improved action with the plaquette
action for gluons, and the SW-clover action with the Wilson action for
quarks.  We find a significant improvement in the hadron spectrum
by improving the quark action, while the gluon improvement is
crucial for a rotationally invariant static potential.  We also explore the
region of light quark masses corresponding to $m_{\rm PS}/m_{\rm V} \geq
0.4$ on a 2.7~fm lattice using the improved gauge and quark action.
A flattening of the potential is not observed up to 2~fm.

\end{abstract}


%////////////////////////////////////////////////////////////////////////////

\section{Introduction}

With the progress over the last few years of quenched simulations of QCD,
it has become increasingly clear that the quenched hadron spectrum shows 
deviations from the experiment if examined at a precision better than 
5--10\%. 
For light hadrons the first indication was that the strange quark mass
cannot be  
set consistently from pseudo scalar and vector meson channels in quenched QCD
\cite{UKQCDJ,LosAlamos96,UKQCDKphi}.
For heavy quark systems calculations both with relativistic
\cite{Fermilab} and non-relativistic \cite{NRQCD} quark actions have shown 
that the fine structure of quarkonium spectra 
can not be reproduced on quenched gluon configurations. 
Most recently an extensive calculation by the CP-PACS collaboration 
found a systematic departure of both the light meson and baryon 
spectra from experiment \cite{quench.CPPACS}. 
These results raise the question as to whether the discrepancies can be 
accounted for by the inclusion of dynamical sea quarks.  
It is therefore timely to study more thoroughly the effects of full QCD in
order to answer this question. 

Full QCD simulations are, however, computationally much more expensive than
those of quenched QCD. Simple scaling estimates coupled with   
past experience place a hundred-fold 
or more increase in the amount of computations for full QCD 
compared to that of quenched QCD with current algorithms. 
Since $32^3\!\times\!64$ is a typical maximal lattice size 
for quenched QCD which can be 
simulated with high statistics on computers with a speed in the 10 
GFLOPS range \cite{LosAlamos96,GF11}, reliable full QCD results are difficult 
to obtain on lattice sizes exceeding $32^3\!\times\!64$
even with TFLOPS-class computers such as CP-PACS \cite{CPPACS-machine}
and QCDSP \cite{QCDSP-machine}. 
Recalling that a physical lattice size of 
$L\approx 2.5$--3.0~fm is needed to avoid finite-size effects
\cite{GF11,FukugitaAoki,MILC-fullKS},  
the smallest lattice spacing one can reasonably reach at present is 
therefore  $a^{-1}\approx 2$~GeV.  Hence lattice discretization errors 
have to be controlled through simulations carried out at inverse lattice 
spacings below this value, {\it e.g.} in the range $a^{-1}\approx 1-2$~GeV.
It is, however, known that with the standard plaquette 
and Wilson quark actions 
discretization errors are already of order 10\% even for 
$a^{-1}\approx 2$~GeV.
These observations suggest the use of improved actions for simulations of
full QCD. 

Studies of improved actions have been widely pursued in the last few 
years. Detailed tests of improvement for the hadron spectrum, 
however, have been carried out mostly within quenched QCD 
\cite{UKQCD-IMPquench,cornell-quench,SCRI-IMPquench,bock,QCDSF,MILC-IMPquench,torvergata,sapienza} 
with only a few full QCD attempts
\cite{SCRI-IMPfull,MILC-IMPfull,UKQCD-IMPfull}. 
In particular, a systematic investigation 
of how gauge and quark action improvement, taken separately,
affects light hadron observables has not been carried out in full QCD.
Prior to embarking on a large scale simulation, 
we examine this question as the first subject of the full QCD program 
on the CP-PACS computer. 

For a systematic comparison of action improvement 
we employ four possible types of action combinations, 
the standard plaquette or a renormalization-group improved 
action \cite{RGIA} for the gauge part, 
and the standard Wilson or the improvement of Sheikholeslami and Wohlert 
\cite{clover} for the quark part.
Since effects of improvement are clearer to discern at coarser lattice 
spacings, we carry out simulations at an inverse lattice spacing of 
$a^{-1} \approx 1$~GeV with quark masses in the range corresponding to 
$m_{\rm PS}/m_{\rm V} \approx 0.7$--0.9.  
Results for the four action combinations are used 
for comparative tests of improvement on the 
light hadron spectrum and the static quark potential.

Another limiting factor for full QCD simulations is how 
close one can approach the chiral limit with present computing power.  
To investigate this question we take the action in which 
both gauge and quark parts are improved, and carry out simulations 
down to a quark mass corresponding to $m_{\rm PS}/m_{\rm V} \approx 0.4$. 
In addition to exploring the chiral behavior of hadron masses, this simulation 
allows an examination of signs of string breaking in the static 
quark-antiquark potential.

In this article we present results of our study on the two questions 
discussed above, expounding on the preliminary accounts reported in 
Refs.~\cite{compar-Tsukuba97,compar-Lattice97}.  
We begin with discussions on our choice of actions for 
our comparative studies in Sec.~\ref{sec:action}.
Details of the full QCD configuration
generation procedure and measurements of hadron masses and potential 
are described in Sec.~\ref{sec:simulations}. 
Results for the hadron masses are discussed in Sec.~\ref{sec:spectrum} 
where, after a description of the chiral extrapolation or interpolation of
our data, we examine the effects of action improvement for the scaling 
behavior of hadron mass ratios. 
In Sec.~\ref{sec:potential} we turn to discuss the static potential. 
The influence of action improvement on 
the restoration of rotational symmetry of the potential is examined, 
and the consistency of the lattice spacing determined from the
vector meson mass and the string tension is discussed.
In Sec.~\ref{sec:TXL} we report on our effort to approach the chiral
limit, where our attempt to observe a flattening of the potential 
at large distances due to string breaking is also presented. We end with a 
brief conclusion in Sec.~\ref{sec:conclusion}. 
Detailed numerical results on run performances, hadron masses and string
tensions are collected at the end in Appendices~\ref{appendix:param},
\ref{appendix:mass} and \ref{appendix:string}.

%////////////////////////////////////////////////////////////////////////////

\section{Choice of action}
\label{sec:action}

The discretization error of the standard plaquette gauge action is 
$O(a^2)$ while that of the Wilson quark action is $O(a)$. 
In principle one would only need to improve the quark action to
the same order as the gauge action. 
On the other hand, violations of rotational invariance have been found to
be strong for the plaquette gauge action at coarse lattice spacings
\cite{cornell-tad,PA4RotSym}.
Hence improving the gauge action is still 
advantageous for coarse lattices. In this spirit we
employ (besides the standard actions) improved actions both 
in the gauge and quark sectors in the forms specified below.

Let us denote the standard plaquette gauge action by \P .  
Improving this action requires the addition of Wilson loops with a
perimeter of six links or more. The number, the precise form and the
coefficients of the added terms differ depending on the principle one
follows for the improvement \cite{improve-overview}.
In this study we test the action determined by an approximate block-spin 
renormalization group analysis of Wilson loops, denoted by \R\ in 
the pursuant, which is given by \cite{RGIA}
\begin{equation}
S_g^{\rm \bf R} = {\beta \over 6}\left(c_0 \sum W_{1\times 1}
               + c_1 \sum W_{1\times 2}\right),
\label{eq:Raction} 
\end{equation}
where the $1\times 2$ rectangular shaped Wilson loop $W_{1\times 2}$
has the coefficient $c_1=-0.331$ and from the normalization 
condition defining the bare coupling $\beta=6/g_0^2$ follows
$c_0=1-8c_1=3.648$.  

The discretization error of the \R\ action is still $O(a^2)$.  The
coefficients of $O(a^2)$ terms in physical quantities, however, are
expected to be reduced from those of the plaquette action.  Indeed, the
quenched static quark potential calculated with this action was found to
exhibit good 
rotational symmetry and scaling already at $a^{-1}\approx 1$~GeV
\cite{kaneko.Tcs}, and so does the scaling of the ratio $T_c/\sqrt{\sigma}$
of the critical temperature of the pure gauge deconfining phase transition
and the string tension $\sigma$ \cite{kaneko.Tcs}.  The degree of
improvement is similar to those observed for tadpole-improved and fixed
point actions \cite{cornell-tad,PA4RotSym}.

To improve the quark action we adopt the clover improvement
proposed by Sheikholeslami and Wohlert \cite{clover}, denoted by \C\ in
the following and defined by
\begin{equation}
D_{xy}^{\rm \bf C}  =  D_{xy}^{\rm \bf W} 
 - \, \delta_{xy} c_{\rm SW} K \sum_{\mu < \nu} 
         \sigma_{\mu\nu} F_{\mu\nu} ,
\label{eq:Caction}
\end{equation}
where $D_{xy}^{\rm \bf W}$ is the standard Wilson quark matrix given by 
\begin{equation}
D_{xy}^{\rm \bf W} = \delta_{xy} 
- K \sum_\mu \{(1-\gamma_\mu)U_{x,\mu} \delta_{x+\hat\mu,y} 
      + (1+\gamma_\mu)U_{x,\mu}^{\dag} \delta_{x,y+\hat\mu} \}
\label{eq:Waction}
\end{equation}
and $F_{\mu\nu}$ is the lattice discretization of the field strength, 

\begin{equation}
F_{\mu\nu} = \frac{1}{8i} (f_{\mu\nu} - f_{\mu\nu}^{\dag}),
\label{eq:CloverLeaf}
\end{equation}
where $f_{\mu\nu}$ is the standard clover-shaped combination of gauge links.

The complete removal of $O(a)$ errors requires a non-perturbative tuning 
of the clover coefficient $c_{\rm SW}$.  This has been carried out 
for the plaquette gauge action in both quenched \cite{alpha,scri-NPR} and 
two-flavor full QCD \cite{alpha-fullqcd}.  
A similar analysis for the \R\ gauge action is yet to be made, however. 
In this study we compare three different choices:
\begin{itemize} 
\item[(a)] 
The tree level value $c_{\rm SW}=1$.
\item[(b)] 
The mean-field (MF) improved value \cite{TadImp} $c_{\rm SW} = P^{-3/4}$
with $P$ the self-consistently determined plaquette average. 
\item[(c)] 
A perturbative mean-field (pMF) improved value 
$c_{\rm SW} = P^{-3/4}$ with the plaquette $P$ calculated in
one-loop perturbation theory. 
For the \R\ gauge action  $P=1-0.8412\cdot\beta^{-1}$ \cite{RGIA}.
\end{itemize}
 
For all three choices the leading discretization error in physical
quantities is $O(g_0^2a)$. The magnitude of the coefficients of this term
should be reduced in the cases of (b) and (c) as compared to (a).  The
one-loop value of $c_{\rm SW}$ has been recently reported to be $c_{\rm
SW}=1+0.678(18)/\beta$ \cite{aoki}. This value is close to the pMF value
$c_{\rm SW}^{\rm pMF}=1+0.631/\beta+\cdots$.  We also find that the
one-loop value of $P$ reproduces the measured values from simulations
within 10\% for the \R\ action.  Hence the pMF value of the clover
coefficient is similar to the MF value employed in (b).  The advantage of
the pMF choice is that it does not require a self-consistent tuning of
$c_{\rm SW}$ for each choice of $\beta$ and $K$.

We carry out simulations employing either the plaquette (\P) or 
rectangular action (\R) for gluons, combining it with either 
the Wilson (\W) or clover action (\C) for quarks.

%////////////////////////////////////////////////////////////////////////////

\section{Simulations}
\label{sec:simulations}

\subsection{choice of simulation parameters}

We choose the coupling constant $\beta$ so
that it gives an inverse lattice spacing of $a^{-1} \approx 1$~GeV. 
For each action combination we choose at least two values of $\beta$ to
allow us to interpolate (or extrapolate) to a desired common lattice
spacing.

Simulations are generally carried out at three values of the hopping
parameter $K$ corresponding to $m_{\rm PS}/m_{\rm V} \approx 0.7$--0.9. 
The lattice size employed is $12^3\!\times\!32$. 

In Table~\ref{tab:run-overview} we give an overview 
of the calculations performed for the action comparison.  
Details of the simulation parameters at each run are collated in
Appendix~\ref{appendix:param}.
Our procedure for estimating the critical 
hopping parameter $K_c$, and the physical 
scale of lattice spacing either from the $\rho$ meson mass ($a_\rho$) 
or the string tension ($a_\sigma$) will be discussed in 
Sec.~\ref{subsec:chiralextrap} and Sec.~\ref{subsec:consistency}.  

We take the \RCpmf\ action at $\beta=1.9$ to explore how close one can 
take the calculation towards the chiral limit.  
For this study we employ two lattice sizes
$12^3\!\times\!32$ and $16^3\!\times\!32$. 
In Table~\ref{tab:TXL-overview} we list the main
features of these two runs whereas details can be found in
Appendix~\ref{appendix:param}.

\subsection{configuration generation and matrix inversion}

Simulations are carried out for two flavors of dynamical quarks using
the hybrid Monte Carlo (HMC) algorithm. The integration of molecular
dynamics (MD) equations is made with the standard leapfrog scheme and with a
step size $\Delta\tau$ chosen to yield an acceptance ratio of 70--90\% 
for trajectories of unit length. The
actual values chosen for $\Delta\tau$ in each case and the measured 
acceptance are given in Appendix~\ref{appendix:param}.

For the inversion of the fermion matrix we employed the minimal residue (MR)
algorithm for our early simulations but
switched later to BiCGStab\cite{BiCGStab}. In both cases we use an even-odd
preconditioning of the quark matrix $D$. $D$
can be decomposed into
\begin{equation}
D(K) = M  - K (D_{eo}+D_{oe}),
\end{equation}
where $M$ is only defined on single sites and the remaining 
connects neighboring sites. For the Wilson quark action $M$ is a unit
matrix, whereas for the clover action it is non-trivial in color and Dirac
space. The even-odd preconditioning consists of solving the equation
$AG_e = B_e'$ where $A = 1 - K^2 M_e^{-1} D_{eo} M_o^{-1} D_{oe}$ and $B_e'
= M_e^{-1} \left ( B_e + K D_{eo} M_o^{-1} B_o \right )$ instead of the
equation $D(K)G = B$.  As an initial guess for the solution vector on even
sites, the right-hand-side vector $G_e = B_e'$ is used. The preconditioning
requires 
the inversion of the local matrix $M$, which is trivial for the Wilson
quark action. For the clover quark action we precalculate $M^{-1}$ and
store it before the solver starts.

As a stopping condition for the matrix inversion during the fermionic force
evaluation we generally use, on the $12^3\!\times\!32$ lattice,  the criterion
\begin{equation}
r_1=||DG-B||^2 \leq 10^{-10}
\label{eq:stopp} 
\end{equation}
which we found to be approximately equivalent to the condition 
\begin{equation}
r_2=||DG-B||/||G|| \leq 10^{-8}. 
\label{eq:stopp2} 
\end{equation}
The actual stopping conditions chosen for
each run and the number of iterations needed to reach this condition are
listed in Appendix~\ref{appendix:param}.
For the evaluation of the Hamiltonian we choose stricter stopping criteria
for $r_1$ between $10^{-14}$ and $10^{-18}$.  

A necessary condition for the validity of the HMC algorithm is the
reversibility of the MD evolution \cite{MD-revers}.  
The CP-PACS computer, on which the
present work is made, employs 64 bit arithmetic for floating point
operations.  Flipping the sign of momenta after a unit trajectory, with the
stopping condition (\ref{eq:stopp2}) above, we
checked that, (i) the
gauge link and conjugate momenta variables return to the starting values
within a relative error of less than $10^{-7}$ on the average, and (ii)
the relative error in the evaluation of the Hamiltonian is less than
$10^{-10}$ (absolute error better than $10^{-4}$ for the $16^3\!\times\!32$
lattice where the check was made) so that the effects in the accept/reject
procedure are far below the level of statistical fluctuations.
 
At each simulated parameter we first run for 100--200 HMC trajectories of
unit length for thermalization and then generate 500--1500 trajectories for 
measurements. Hadron propagators are measured on configurations separated
by 5 trajectories.  The static quark potential is measured on a subset of 
the configurations separated by either 5 or 10 trajectories.  
The detailed numbers are again given in Appendix~\ref{appendix:param}.

\subsection{hadron mass measurement}

We calculate quark propagators for the hopping parameter equal to that 
for the dynamical quarks used in the configuration generation. 
Two quark propagators are prepared for each configuration, one with 
the point source and the other with an exponentially smeared source 
with the smearing function $\psi(r) = A\exp(-Br)$.  
For the latter we fix the gauge configuration to the Coulomb gauge.  
The choice of the smearing parameters $A$ and $B$ is guided by 
previous quenched results for the pion wave function \cite{JLQCD-smearing}, 
readjusted by hand so that hadron effective masses reach a plateau as soon 
as possible.

\begin{figure*}[p]
\vspace{3mm}
\centerline{
\epsfxsize=13cm \epsfbox{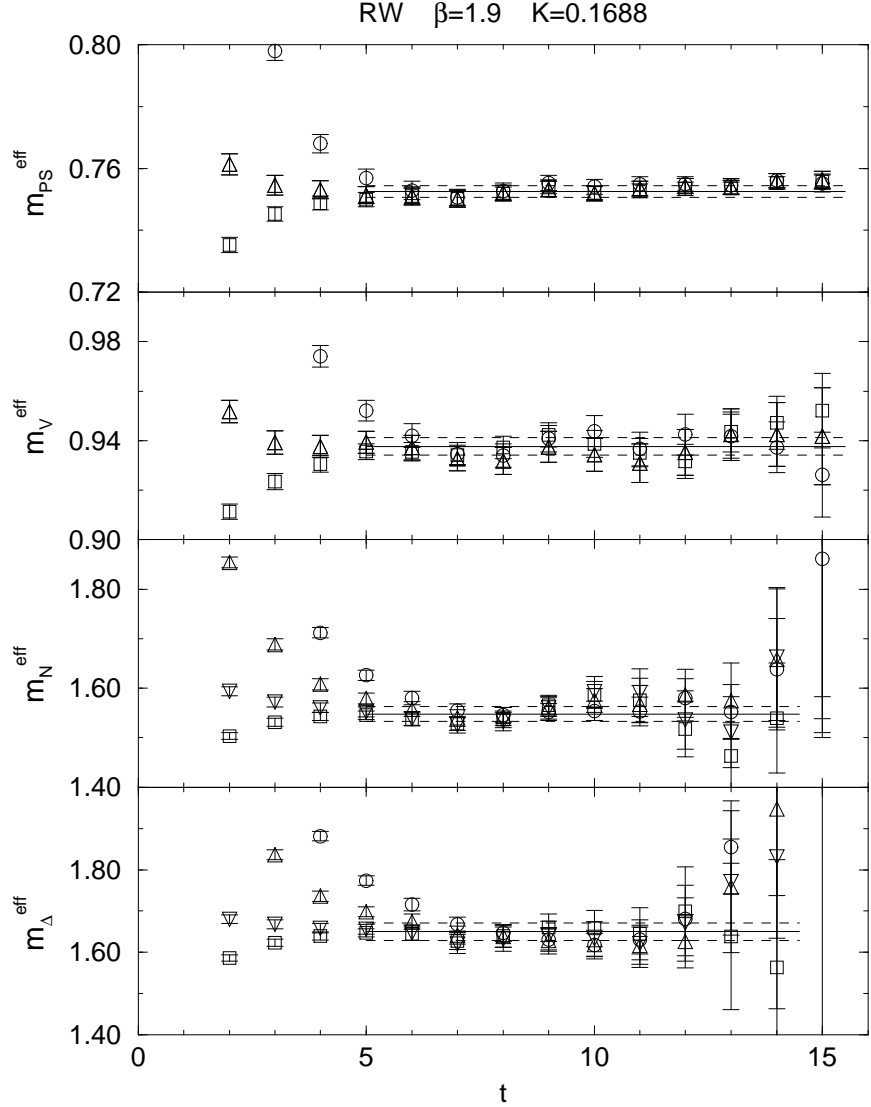}
}
\vspace{-5mm}
\caption{Example of effective mass plots for pseudo scalar, vector, nucleon
and $\Delta$ on a $12^3\!\times\!32$ lattice. Circles are effective masses
where all quark propagators have  
point sources (PP or PPP). For squares all quark propagators have
smeared sources (SS or SSS) and triangles are for mixed combinations of
sources (PS, PPS or PSS). Solid lines denote the results from mass fits to
SS or SSS correlators. Dashed lines show the one standard deviation error
band determined by jackknife analysis.}
\label{fig:eff-mass}
\vspace{-4mm}
\end{figure*}

Hadron propagators are constructed by combining quark propagators 
for the point (P) or the smeared (S) sources in various ways, but always 
adopting the point sink. For example, PS represents a meson propagator 
calculated with the point source for quark and the smeared source for 
antiquark. 
In Fig.~\ref{fig:eff-mass} we show a typical
example of effective masses for a variety of source combinations. 

In most cases the effective masses for the SS (SSS for baryons)
propagators comes 
from below, show the best plateau behavior,  and have the smallest 
statistical errors estimated with the jackknife procedure. 
We therefore determine hadron masses with a fit to SS (SSS) hadron 
propagators, supplementing results for other source combinations 
as a guide in choosing the fit range.

Hadron masses are extracted from propagators by employing a single hyperbolic
cosine fit for mesons and a single exponential fit for baryons. We use
uncorrelated fits and determine the error with the jackknife method. 
While our runs of at most 1500 HMC trajectories are not really long enough 
to carry out detailed autocorrelation analysis, examining the bin size 
dependence of the estimated error indicates 
a bin size of 5 configurations or 25 HMC trajectories to be a reasonable 
choice, which we adopt for all of our error analyses. 

The hadron mass results for all our runs are collected in
Appendix~\ref{appendix:mass}.

\subsection{potential measurement}
\label{subsec:pot.meas}

We measure Wilson loops $W(r,t)$  both in the on- and off-axis directions in 
space.  The spatial paths of $W(r,t)$ are formed by
connecting one of the following spatial vectors repeatedly,
\begin{equation}
(1,0,0),\:  (1,1,0),\:  (2,1,0),\: (1,1,1),\: (2,1,1),\: (2,2,1) .
\end{equation}
We measure $W(r,t)$ up to $r \le 6$ and $t\le8$ on the $12^3\!\times\!32$ 
lattice, while we enlarge the largest spatial size
to $r \le 4\sqrt{3}$ on the $16^3\!\times\!32$ lattice  
in order to investigate the large distance behavior of the potential. 
The smearing procedure of Ref.~\cite{Wsmearing} is applied to 
the link variables, up to 6 times on the $12^3\!\times\!32$ lattice 
and up to 8 times on the $16^3\!\times\!32$ lattice, respectively.
The Wilson loop is measured at every smearing step in order  
to choose the optimal smearing number for each value of $r$.

We extract the potential $V(r)$ and the overlap function $C(r)$
by a fully correlated fit of the Wilson loop 
to the form 
\begin{equation}
W(r,t) = C(r)\ {\rm exp}\left[ -V(r)t \right] .
\label{eqn:fit1}
\end{equation}
The optimum smearing number at each $r$ is determined 
by the condition that the overlap $C(r)$ takes the largest value 
smaller than 1.

\begin{figure*}[p]
\vspace{-5mm}
\centerline{
              \epsfxsize=8cm \epsfbox{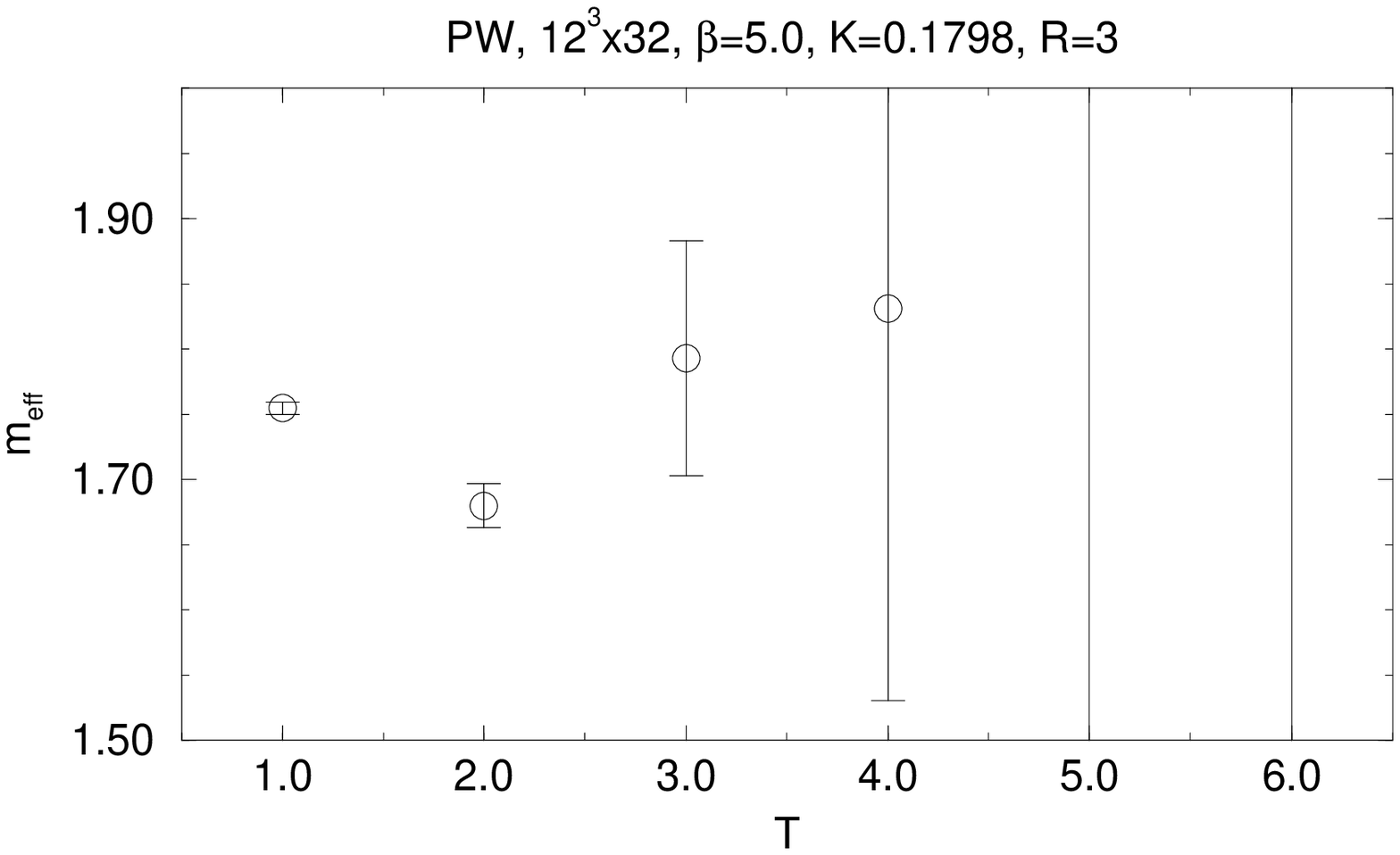}
\hspace{-8mm} \epsfxsize=8cm \epsfbox{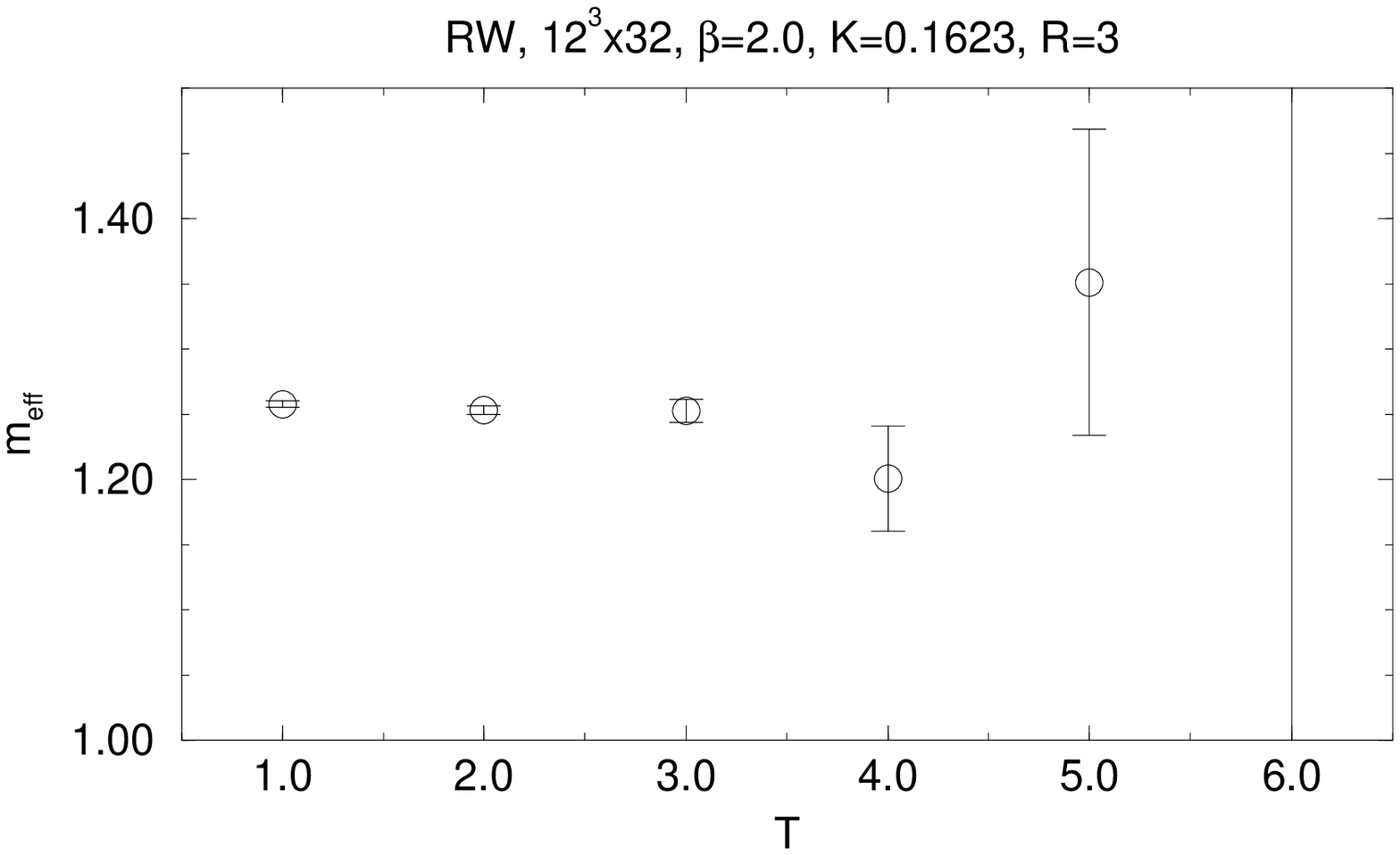}
}
\vspace{-20mm}
\centerline{
              \epsfxsize=8cm \epsfbox{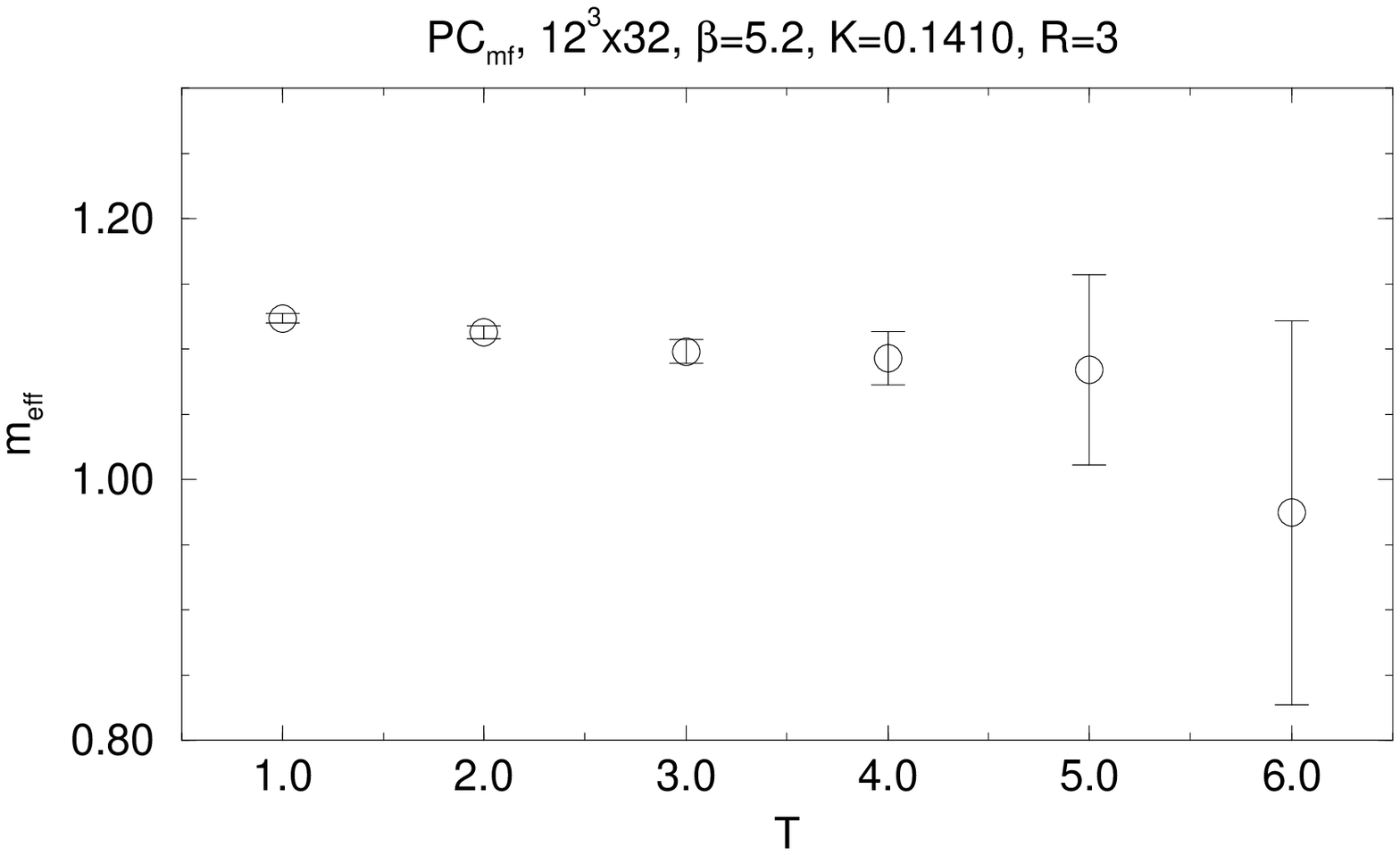}
\hspace{-8mm} \epsfxsize=8cm \epsfbox{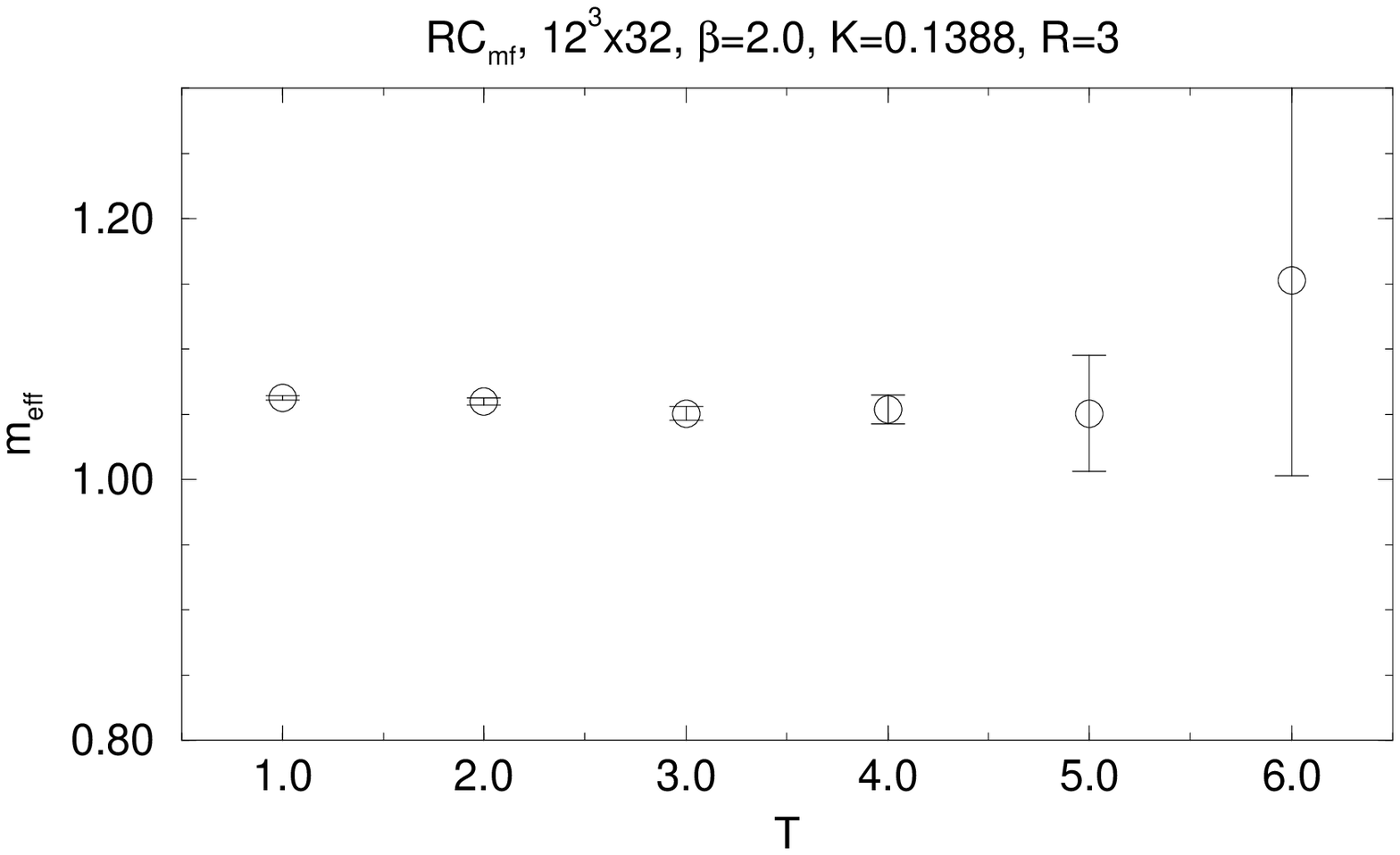}
}
\vspace{-20mm}
\caption{
Effective masses of the static quark potential for the optimum smearing
at $r=3a$ for four action combinations.
}
\label{fig:em}
\vspace{-4mm}
\end{figure*}

Typical results for the effective mass defined by
\begin{equation}
m_{\rm eff} = {\rm ln}\left[ W(r,t)/W(r,t+1) \right],
\label{eqn:em}
\end{equation}
are shown in Fig.~\ref{fig:em}. We find that noise generally dominates 
over the signal for $t > 4$. 
Thus we set the upper limit of the fitting range to $t_{\rm max} = 4$. 
Since choosing the lower limit $t_{\rm min} = 1$ leads to an increase of 
${\chi}^2/{\rm d.o.f}$ by 3--10 times compared to the choice $t_{\rm min}=2$ 
for most values of $r$ and simulation parameters, 
we fix the fitting range to be $t=2$--4.

The statistical  error of $V(r)$ is estimated 
by the jackknife method.  We find that a bin size of 30 HMC trajectories
is generally sufficient to ensure stability of errors against bin size. 
We therefore adopt this bin size for all of our error estimates with 
potential data.

%////////////////////////////////////////////////////////////////////////////

\section{Hadron spectrum}
\label{sec:spectrum}

\subsection{chiral fits}
\label{subsec:chiralextrap}

A basic parameter characterizing the chiral behavior of hadron masses
is the critical hopping parameter
$K_c$ at which the pseudo scalar meson mass $m_{\rm PS}a$ vanishes.  
Results for $(m_{\rm PS}a)^2$ exhibit deviations from
a linear function in $1/K$, and hence we extract $K_c$ by assuming
\begin{equation}
(m_{\rm PS}a)^2 =   B_{\rm PS}\left(\frac{1}{K}-\frac{1}{K_c}\right)
                  + C_{\rm PS}\left(\frac{1}{K}-\frac{1}{K_c}\right)^2. 
\end{equation}
The fitted values of the critical hopping parameter are listed in 
Table~\ref{tab:run-overview} and \ref{tab:TXL-overview}. 

Another important parameter is the vector meson mass $m_{\rm V}a$ in the
chiral limit $m_{\rm PS}a=0$, which allows us to set the physical lattice 
spacing. 
We determine this quantity by a chiral fit of the vector meson mass 
in terms of the pseudo scalar meson mass, both of which are measured 
quantities. Our results for this relation show curvature (see
Fig.~\ref{fig:chiralRC19} in Sec.~\ref{sec:small_mass} for an example), 
and hence for the fitting function we employ 
\begin{equation}
m_{\rm V}a=A_{\rm V}+B_{\rm V}(m_{\rm PS}a)^2+C_{\rm V}(m_{\rm PS}a)^3,
\label{eq:cubic}
\end{equation}
where the cubic term is inspired by chiral perturbation theory. 

A practical problem with this fit is that we have only three data points
for most of our runs. We
estimate systematic uncertainties in the extrapolation by repeating the fit 
without the cubic term to the two points of data for lighter quark masses. 
Results for the vector meson mass in the chiral limit, translated into 
the lattice spacing through $a_\rho=A_{\rm V}/768{\rm MeV}$, are listed in 
Table~\ref{tab:run-overview} and \ref{tab:TXL-overview}.  

Results for the nucleon and $\Delta$ also show curvature in terms of 
$m_{\rm PS}a$. We therefore fit them employing a cubic polynomial 
without the linear term (\ref{eq:cubic}) as for the vector meson mass. 

\subsection{scaling of mass ratios}

\begin{figure*}[p]
\vspace{-1mm}
\centerline{
              \epsfxsize=8cm \epsfbox{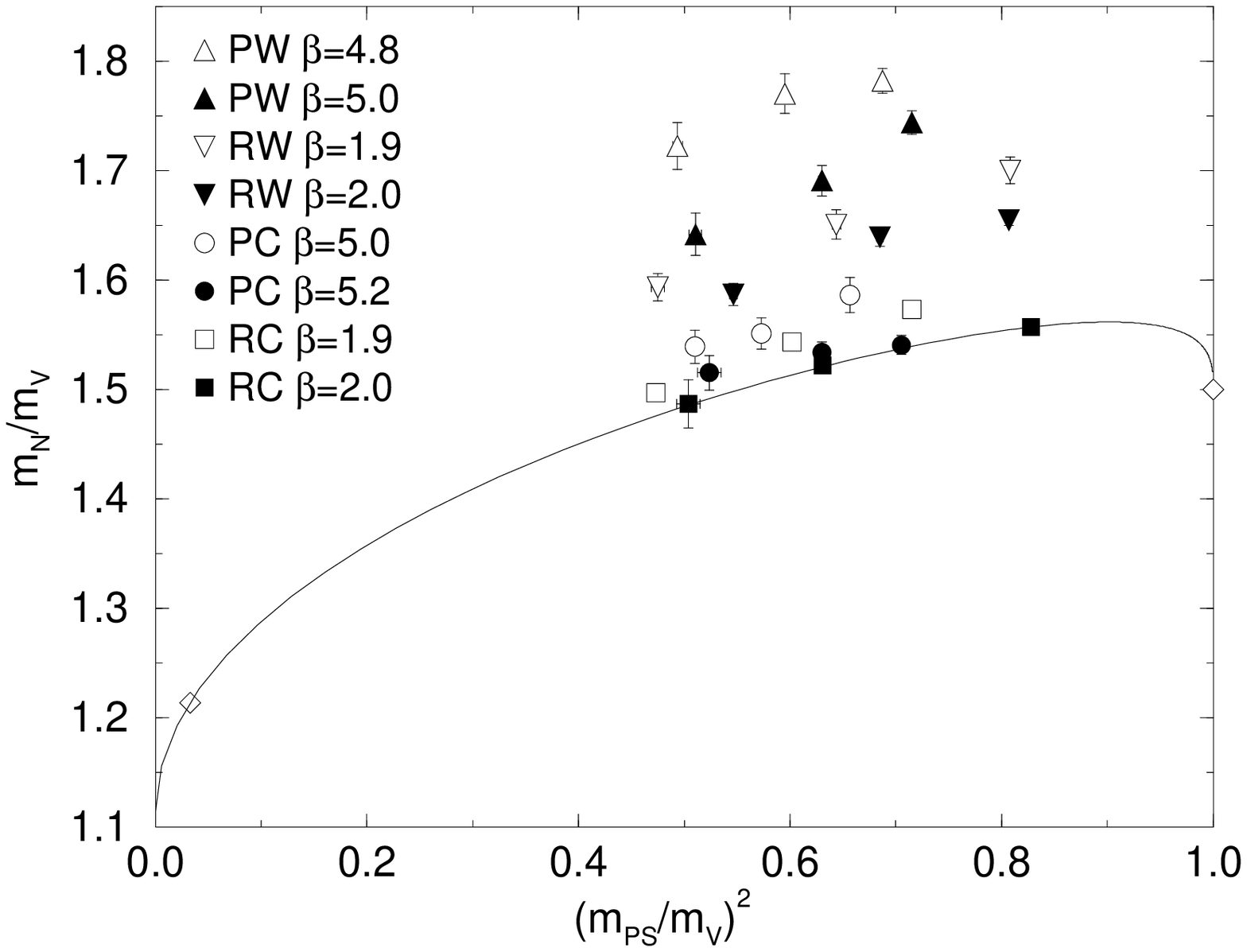}
\hspace{-8mm} \epsfxsize=8cm \epsfbox{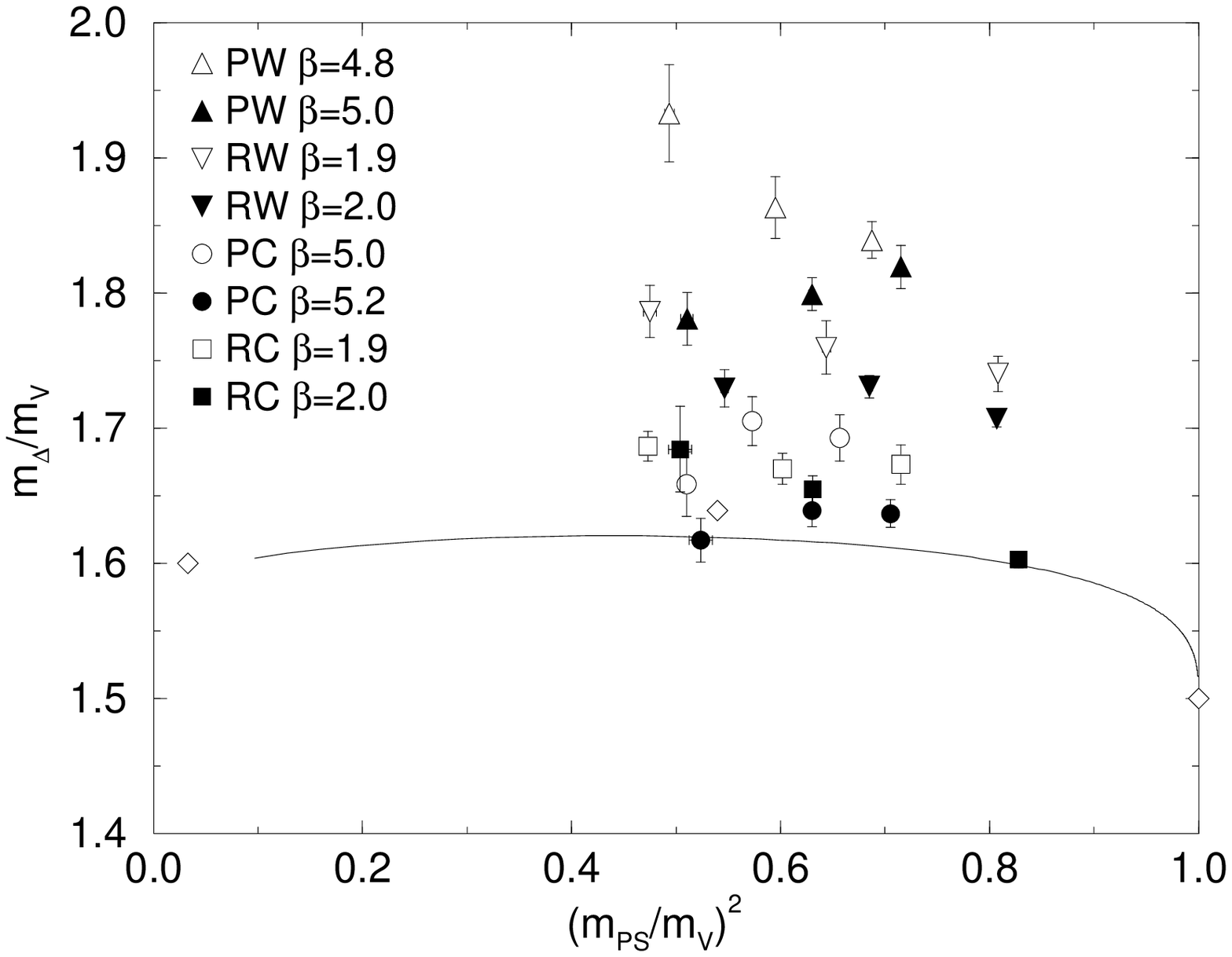}
}
\vspace{-40mm}
\caption{$m_{\rm N}/m_{\rm V}$ and $m_{\Delta}/m_{\rm V}$ as function of
$(m_{\rm PS}/m_{\rm V})^2$ for four combinations of the action.
Diamonds are experimental points corresponding to ${\rm
N}(940)/\rho(770)$, $\Delta(1232)/\rho(770)$ and $\Omega(1672)/\phi(1020)$.
}
\label{fig:ape}
\vspace{-4mm}
\end{figure*}

We show in Fig.~\ref{fig:ape} a compilation of our hadron mass results 
for the four action combinations in terms of the 
mass ratios $m_{\rm N}/m_{\rm V}$ and $m_{\Delta}/m_{\rm V}$ 
as a function of $(m_{\rm PS}/m_{\rm V})^2$.  In order to avoid 
overcluttering of points, we include results for only two values of $\beta$
per action combination. Furthermore, for the \PC\ action combination 
the results with $c_{\rm SW} =$~MF are displayed whereas for the \RC\ action 
results for $c_{\rm SW} =$~pMF are shown. 

We observe two features in this figure. In the first instance, for each
action combination  
the baryon to vector meson  mass ratio decreases as the coupling decreases. 
This is a well-known trend of scaling violation for Wilson-type quark actions.
Secondly, the magnitude of scaling violation, measured by the distance 
from the phenomenological curve (solid line in Fig.~\ref{fig:ape}) 
\cite{ONO} has an order where $\PW > \RW > \PC >\RC$.
In particular the results for the \PC\ and \RC\ cases show a significant 
improvement over those for the \PW\ and \RW\ cases 
in that they lie close 
to the phenomenological curve even though the lattice spacing is as large 
as $a_\rho^{-1}\approx 1$--1.3~GeV 
(see Tables~\ref{tab:run-overview} and \ref{tab:TXL-overview}). 

A point of caution, however, is that the lattice spacings for the data sets 
displayed in Fig.~\ref{fig:ape} do not exactly coincide.  In order to 
disentangle effects associated with action improvement from those of 
a finer lattice spacing for each action, we need to plot results 
at the same lattice spacing.
  
One way to make such a comparison is to take a cross section of 
Fig.~\ref{fig:ape} at a fixed value of $m_{\rm PS}/m_{\rm V}$ and plot 
the resulting value of $m_{\rm N, \Delta}/m_{\rm V}$ as a function 
of $m_{\rm V}a$ at that value of $m_{\rm PS}/m_{\rm V}$.  
This requires an interpolation 
of hadron mass results, for which 
we employ the cubic chiral fits described in Sec.~\ref{subsec:chiralextrap}
and the jackknife method for error estimation. 

\begin{figure*}[p]
\vspace{-1mm}
\centerline{
              \epsfxsize=8cm \epsfbox{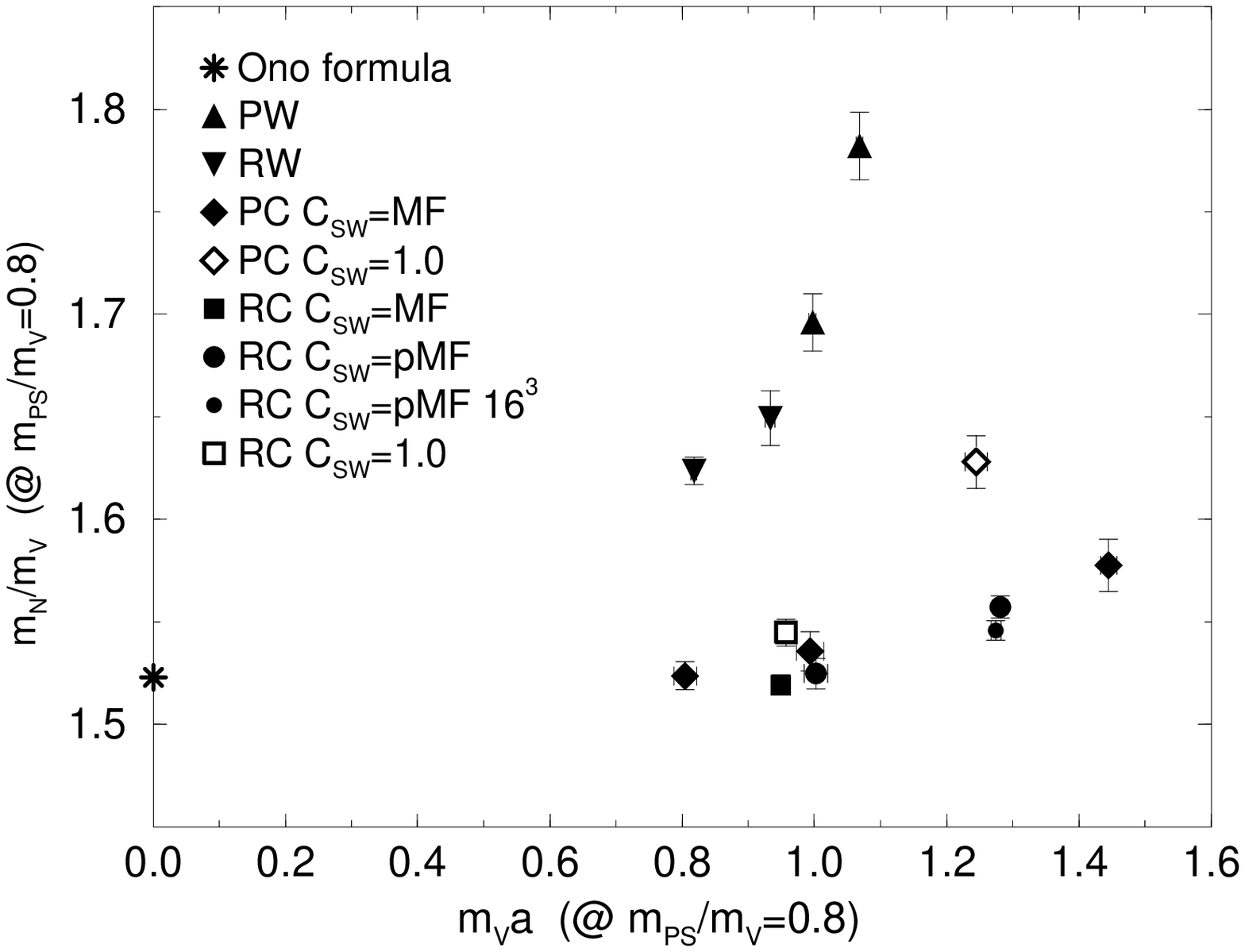}
\hspace{-8mm} \epsfxsize=8cm \epsfbox{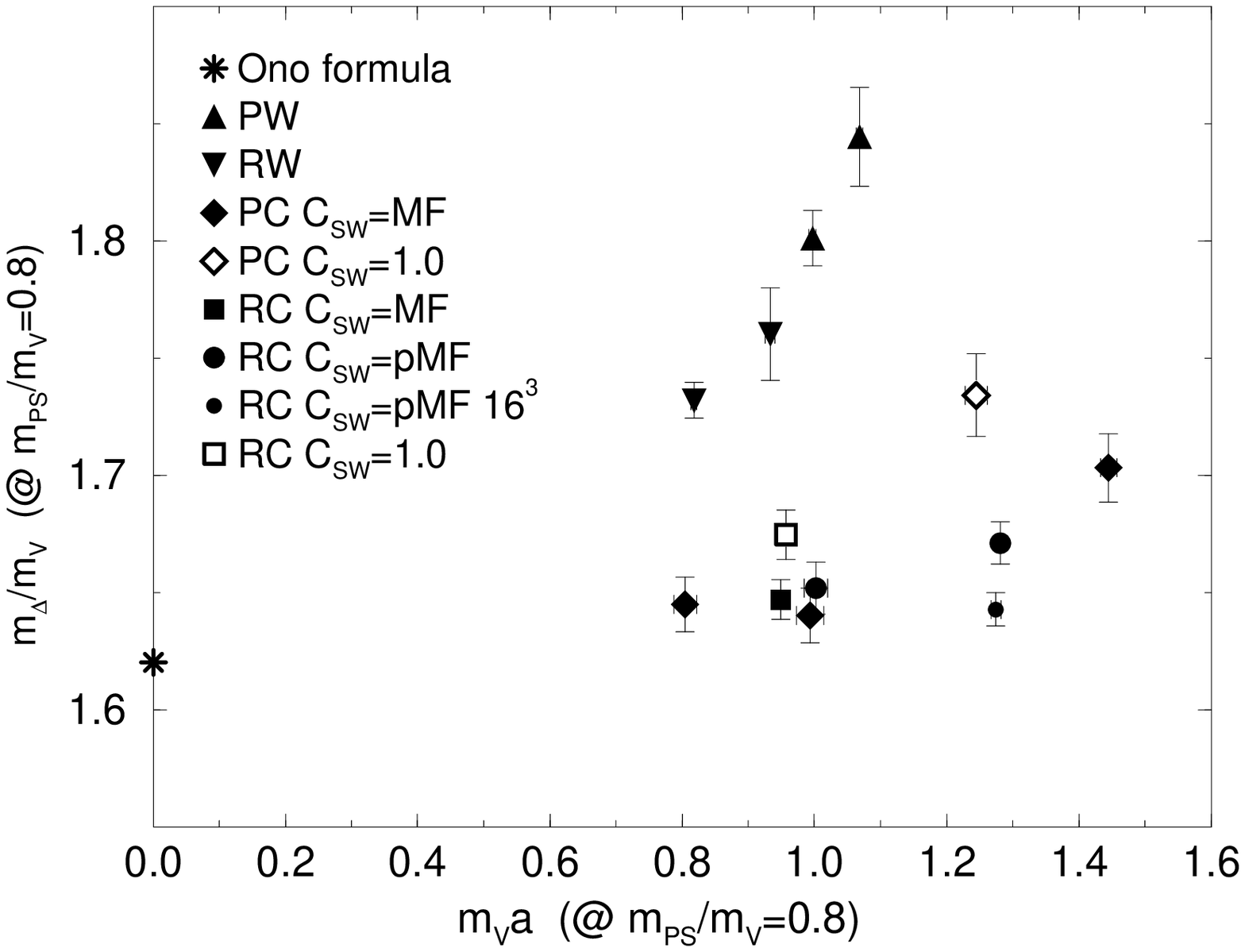}
}
\vspace{-40mm}
\caption{Scaling behavior of $m_{\rm N}/m_{\rm V}$ and $m_{\Delta}/m_{\rm
V}$ at fixed $m_{\rm PS}/m_{\rm V}=0.8$ as function of $m_{\rm V}a$. 
}
\label{fig:scaling-0.8}
\vspace{-4mm}
\end{figure*}

In Fig.~\ref{fig:scaling-0.8} we show results of this analysis 
for $m_{\rm N}/m_{\rm V}$ and $m_{\Delta}/m_{\rm V}$
at $m_{\rm PS}/m_{\rm V}=0.8$.  It is interesting to observe 
that the \PW\ and \RW\ results lie almost on a single curve, while 
the \PC\ and \RC\ results, respectively using the MF and pMF value of
$c_{\rm SW}$,   
fall on a different, much flatter curve.  This clearly shows that the 
improvement of the gauge action has little effect on decreasing the scaling 
violation in the baryon masses.  The improvement is due to the use of the 
clover quark action for the \PC\ and \RC\ cases.  An apparently better 
behavior of \RW\ results in Fig.~\ref{fig:ape} compared to those for the 
\PW\ case is merely an effect of the finer lattice spacing of the former.  

We have commented in Sec.~\ref{sec:action} that the values of $c_{\rm SW}$
for the MF and pMF cases are similar.  This would explain why results
for the \PC\ action with the MF value of $c_{\rm SW}$ and those for the
\RC\ action with the pMF value of $c_{\rm SW}$ lie almost on a single
curve.  For both MF and pMF choices, the magnitude of $c_{\rm SW}$ is
significantly larger than the tree-level value $c_{\rm SW}=1$.  As is shown
in Fig.~\ref{fig:scaling-0.8} with open symbols, the degree of improvement
with the tree-level $c_{\rm SW}$ is substantially less than those for the
MF and pMF choices.

%////////////////////////////////////////////////////////////////////////////

\section{Static quark potential}
\label{sec:potential}

\subsection{restoration of rotational symmetry}

\begin{figure*}[p]
\vspace{-1mm}
\centerline{
\epsfxsize=8cm \epsfbox{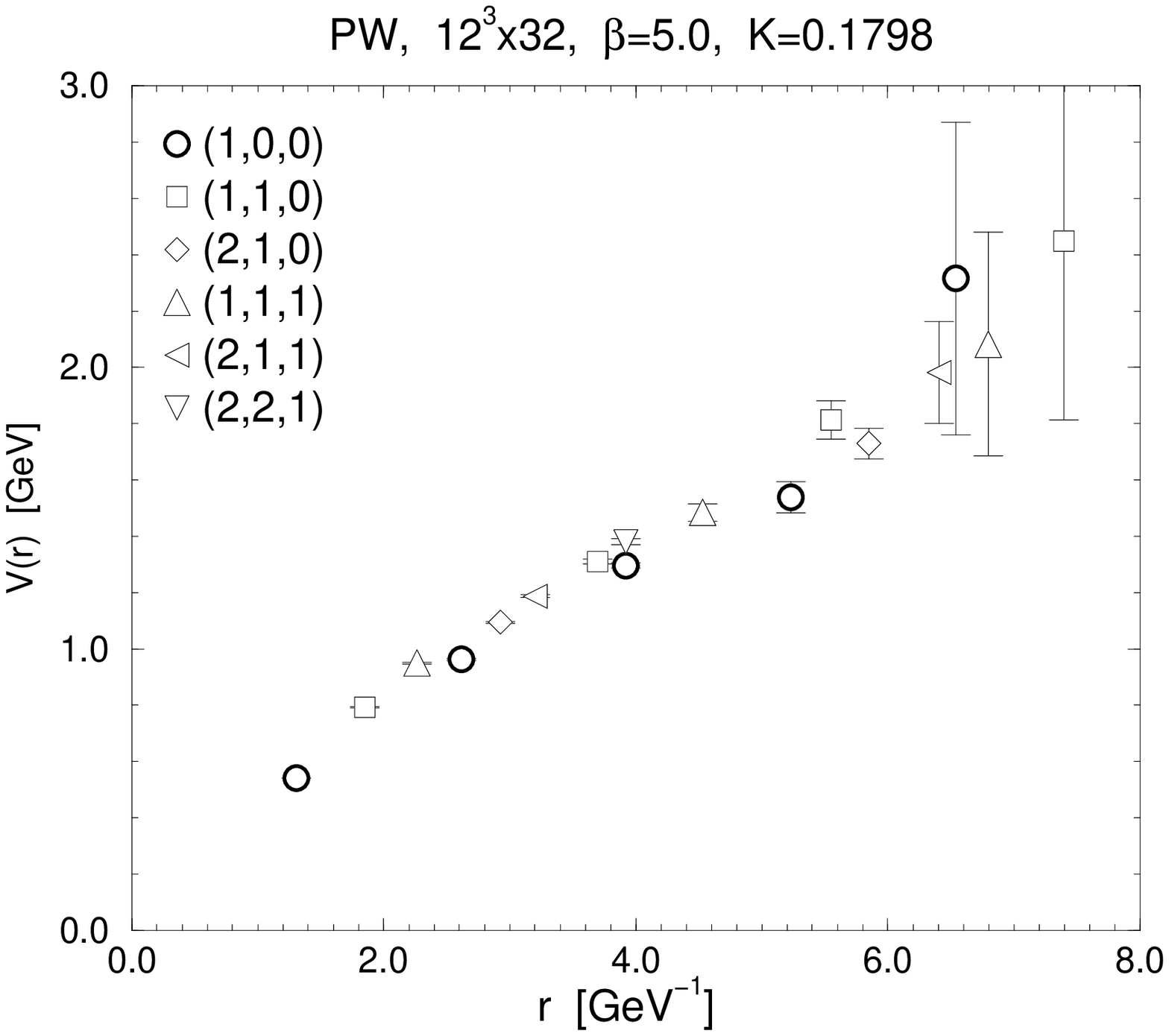}
\epsfxsize=8cm \epsfbox{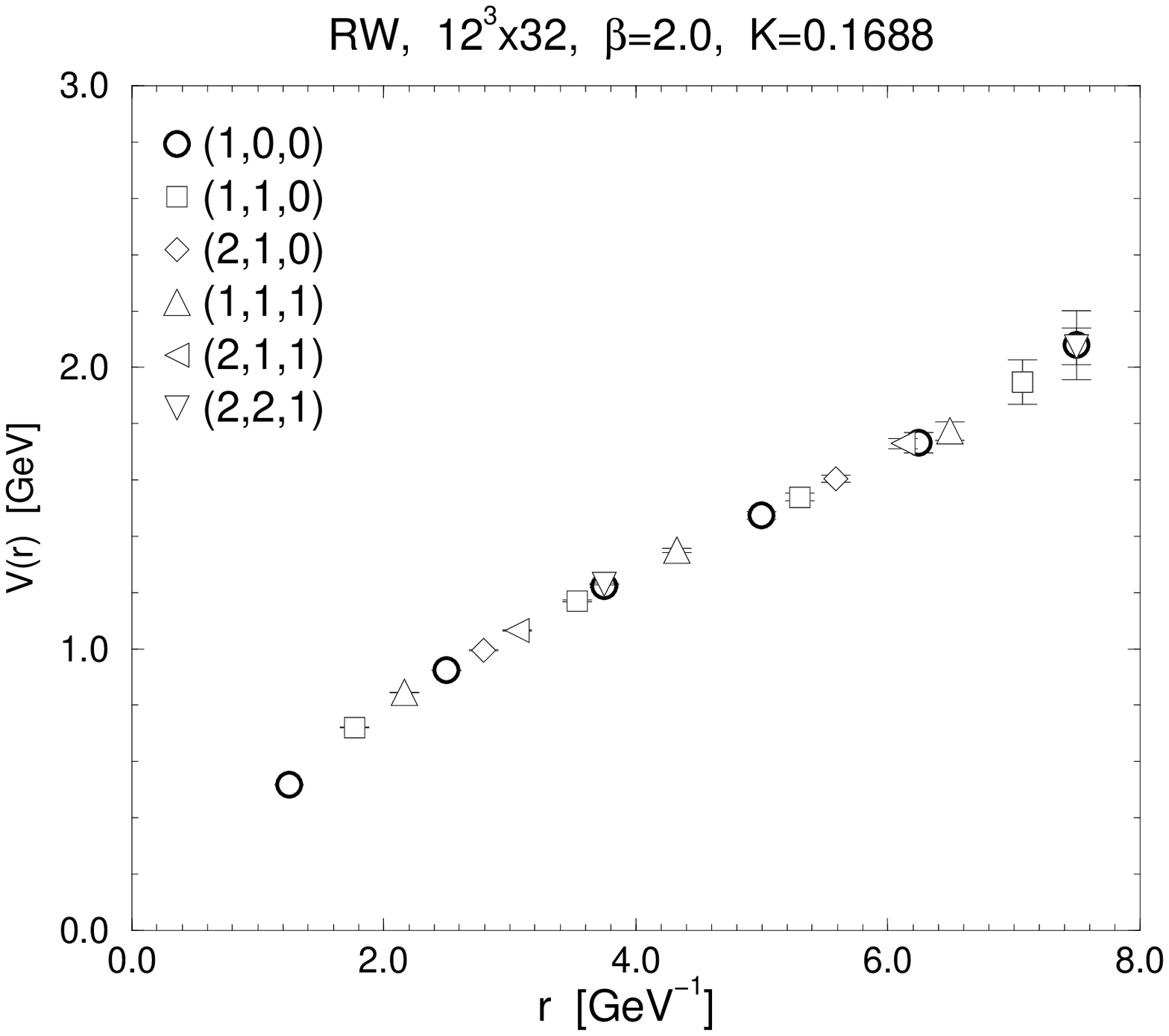}
}
\vspace{2mm}
\centerline{
\epsfxsize=8cm \epsfbox{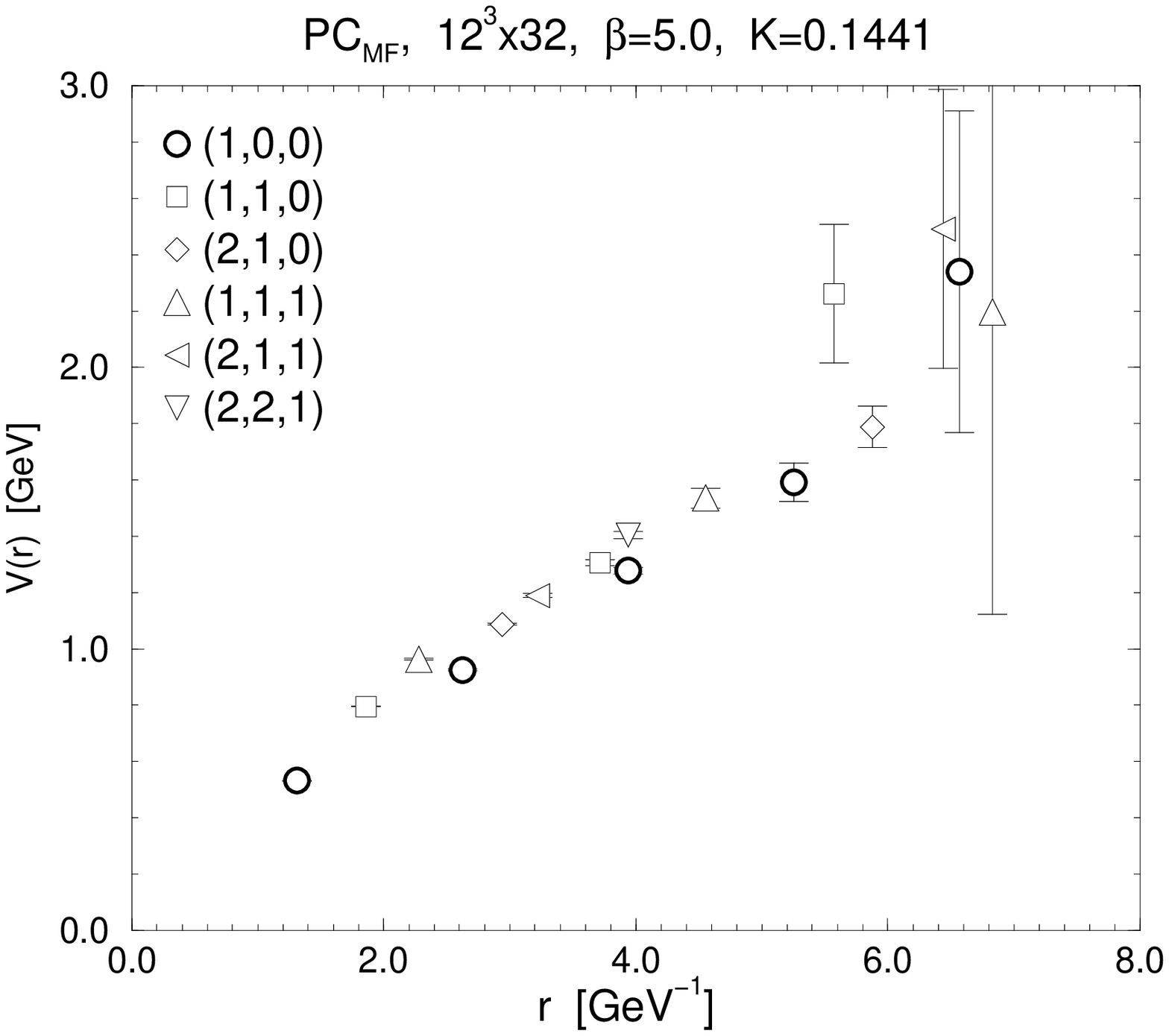}
\epsfxsize=8cm \epsfbox{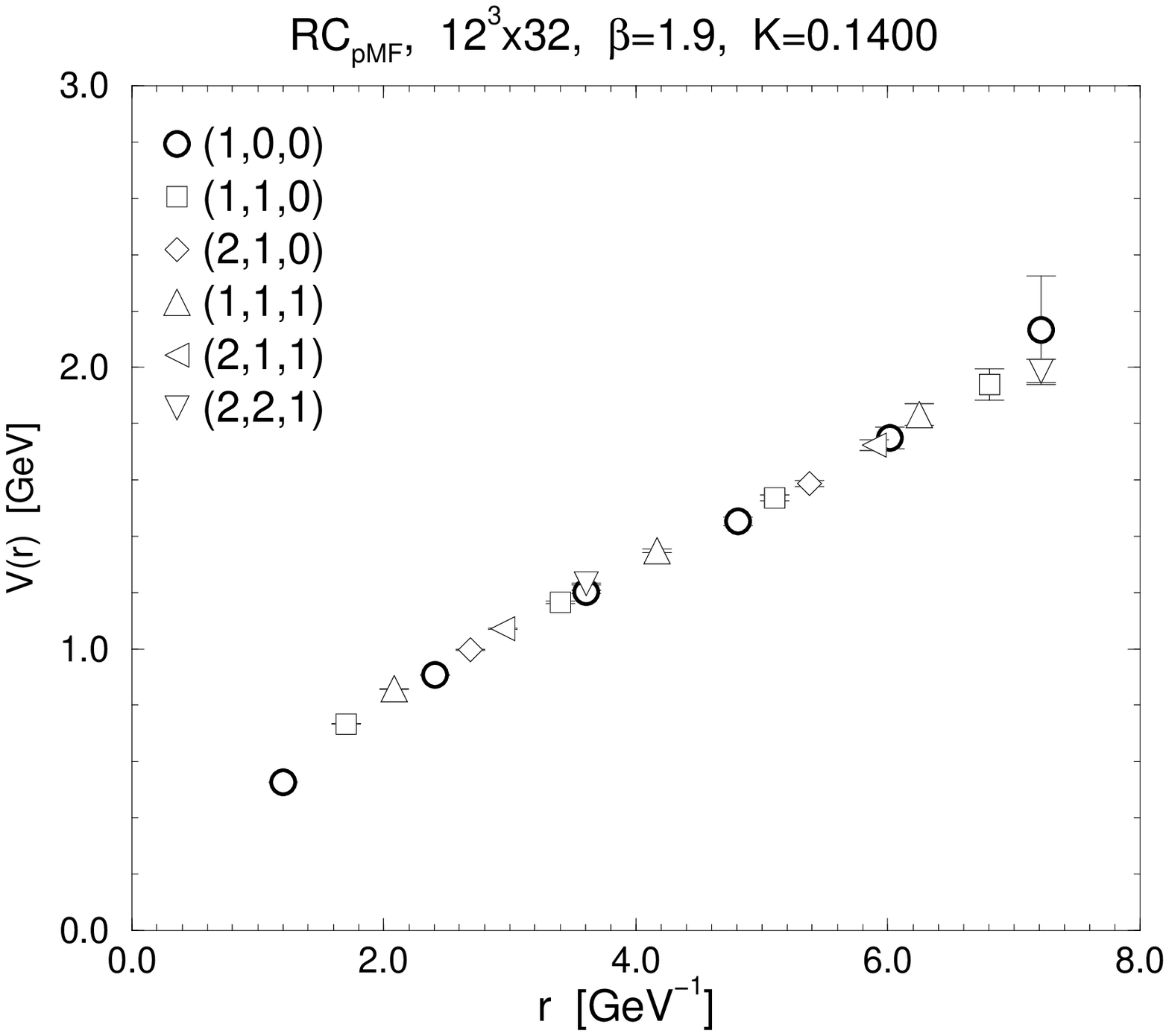}
}
\vspace{1mm}
\caption{
Static quark potential for the four action combinations
at $m_{\rm PS}/m_{\rm V} \simeq 0.8$ on the $12^3\!\times\!32$ lattice
with a lattice spacing $a \approx 1 {\rm GeV^{-1}}$. 
Scales are set by the lattice spacing determined from the string tension.
Different symbols correspond to the potential data measured in different
spatial directions along the vector indicated in the figure.  
}
\label{fig:VvsR.12x32.a010}
\vspace{-4mm}
\end{figure*}

In Fig.~\ref{fig:VvsR.12x32.a010}, we plot our potential 
data for the four action combinations 
at a quark mass corresponding to ${m_{\rm PS}}/{m_{\rm V}} \approx 0.8$
and $a^{-1} \approx 1$~GeV.
We find a sizable violation of rotational symmetry in the \PW\ case 
at this coarse lattice spacing.
Looking at the potential for the \PC\ case, we cannot observe any 
noticeable restoration of the symmetry. 
In contrast,
a remarkable restoration of rotational symmetry is apparent 
in the \RW\ and \RC\ cases.

In order to quantify the violation of rotational symmetry and its 
improvement depending on the action choice, we consider the 
difference between the on-axis and off-axis
potential at a distance $r=3$ defined by 

\begin{equation}
\Delta V = \frac{ V\left( r\!=\!\left(3,0,0\right )\right)
                      -V\left(r\!=\!\left(2,2,1\right)\right) }
                { V\left( r\!=\!\left(3,0,0\right )\right)
                      +V\left( r\!=\!\left(2,2,1\right )\right) }.
\label{eqn:dV}
\end{equation}
We find that the value of $\Delta V$ monotonously decreases as the sea
quark mass decreases for most cases.  We ascribe this trend to the 
fact that one effect of dynamical sea quarks is to renormalize the 
coupling toward a smaller value, and hence reduces violation of rotational 
symmetry. 

\begin{figure*}[p]
\vspace{-1mm}
\centerline{
\epsfxsize=10cm \epsfbox{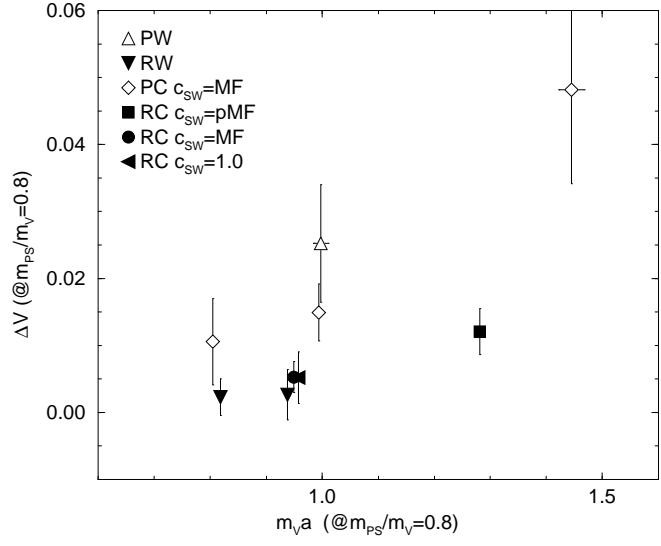}
}
\vspace{1mm}
\caption{
${\Delta}V$ as a function of the vector meson mass $m_{\rm V}a$
at $m_{\rm PS}/m_{\rm V}=0.8$.}
\label{fig:dV.vs.mV}
\vspace{-4mm}
\end{figure*}

In order to make a comparison at the same quark mass,
we estimate $\Delta V$ at $m_{\rm PS}/m_{\rm V}=0.8$
by an interpolation as a linear function of $(m_{\rm PS}a)^2$.
In Fig.~\ref{fig:dV.vs.mV} we plot results for $\Delta V$
obtained in this way  
against the value of $m_{\rm V}a$ at $m_{\rm PS}/m_{\rm V}=0.8$.
This figure confirms the qualitative impression from 
Fig.~\ref{fig:VvsR.12x32.a010}.  Rotational symmetry is badly violated 
for the \PW\ and \PC\ cases, which is significantly improved 
by changing the gauge action as demonstrated by the small values of 
$\Delta V$ for the \RW\ and \RC\ results. 
In contrast the effect of quark action improvement 
on the restoration of rotational symmetry appears to be small. 
This may not be surprising since dynamical quarks affect the static 
potential only indirectly through vacuum polarization effects. 

\subsection{string tension}
\label{sec:sigma}

The static potential in full QCD is expected to flatten at large 
distances due to string breaking.  None of our potential data, which 
typically extends up to the distance of $r\approx 1$~fm, show signs of 
such a behavior, but rather increase linearly. As we discuss in more 
detail in Sec.~\ref{sec:TXL} this is probably due to a poor overlap of 
the Wilson loop operator with the state of a broken string.  
This suggests that we can extract the string tension from 
the present data for the potential $V(r)$ 
by assuming the form 
\begin{equation}
V(r) = V_0 - \frac{\alpha}{r} + \sigma r.
\label{eqn:fit2}
\end{equation}

In practice we find that the Coulomb coefficient $\alpha$ is difficult to 
determine from the fit,  even if we introduce the tree-level correction term 
corresponding to the one lattice gluon exchange diagram.
This may be due to the fact that our potential data taken at 
coarse lattice spacings do not have enough points at short distance 
to constrain the Coulomb term. 
As an alternative we test a two-parameter fitting 
with a fixed Coulomb term coefficient 
${\alpha}_{\rm fixed}=0.1$, 0.125, ..., 0.475, and 0.5, 
using the fitting range $r_{\rm min}$--$r_{\rm max}$
with $r_{\rm min}=1$, $\sqrt{2}$, $\sqrt{3}$ and $r_{\rm max}=5$--6. 
We find that the value of ${\chi}^2/{\rm d.o.f}$ takes its minimum value 
around ${\alpha}_{\rm fixed}=0.3$--$0.4$
for most fitting ranges and simulation parameters.

Based on this result, we extract the string tension by fitting
the potential at large distances, where a linear behavior dominates, 
to the form (\ref{eqn:fit2}) with a fixed Coulomb coefficient 
${\alpha}_{\rm fixed}=0.35$. The shift of the fitted $\sigma$ over 
the range $\alpha=0.3$--$0.4$ is taken into estimates of the systematic error.
 
The result for the string tension $\sigma$ with this two-parameter fit
is quite stable against variations of $r_{\rm max}$. It does depend 
more on $r_{\rm min}$, however.  This leads us to repeat the two-parameter 
fit with ${\alpha}_{\rm fixed}=0.35$ over the interval of $r_{\rm min}$ 
listed in Appendix~\ref{appendix:string},
and determine the central value of $\sigma$ by 
the weighted average of the results over the ranges.
The variance over the ranges are included into the
systematic error of $\sigma$.
We collate the final results for the string tension $\sigma$
in Appendix~\ref{appendix:string}.

\subsection{consistency in lattice spacings}
\label{subsec:consistency}

The scaling violation in the ratio $m_{\rho}/\sqrt{\sigma}$ leads to an
inconsistency in the lattice spacings determined from
the $\rho$ meson mass $a_{\rho}$ and the string tension
$a_{\sigma}$ in the chiral limit.
Thus, examination of this consistency provides another 
test of effectiveness of improved actions.
For the physical value we use $m_\rho=768$~MeV and 
$\sqrt{\sigma}=440$~MeV.
We should note that the latter value is uncertain by about 5--10\% since 
the string tension is not a directly measurable quantity by experiment. 

The chiral extrapolation of the vector meson mass was already discussed in 
Sec.~\ref{subsec:chiralextrap}.  We follow a similar procedure for the 
chiral extrapolation of the string tension. Namely  
we fit results to a form 
\begin{equation}
{\sigma}a^2 = A_{\sigma} + B_{\sigma} {\cdot}(m_{\rm PS}a)^2 
			 + C_{\sigma}{\cdot}(m_{\rm PS}a)^3.
\label{eqn:as.linfit}
\end{equation}
In most cases we find a quadratic ansatz ($C_\sigma=0$) to be 
sufficient, which we then adopt for all data sets. 
Results for the string tension in the chiral limit, converted to the 
physical scale of lattice spacing $a_\sigma$,  are listed in 
Table~\ref{tab:run-overview} and \ref{tab:TXL-overview}. 

\begin{figure*}[b]
\vspace{0mm}
\centerline{
\epsfxsize=8cm \epsfbox{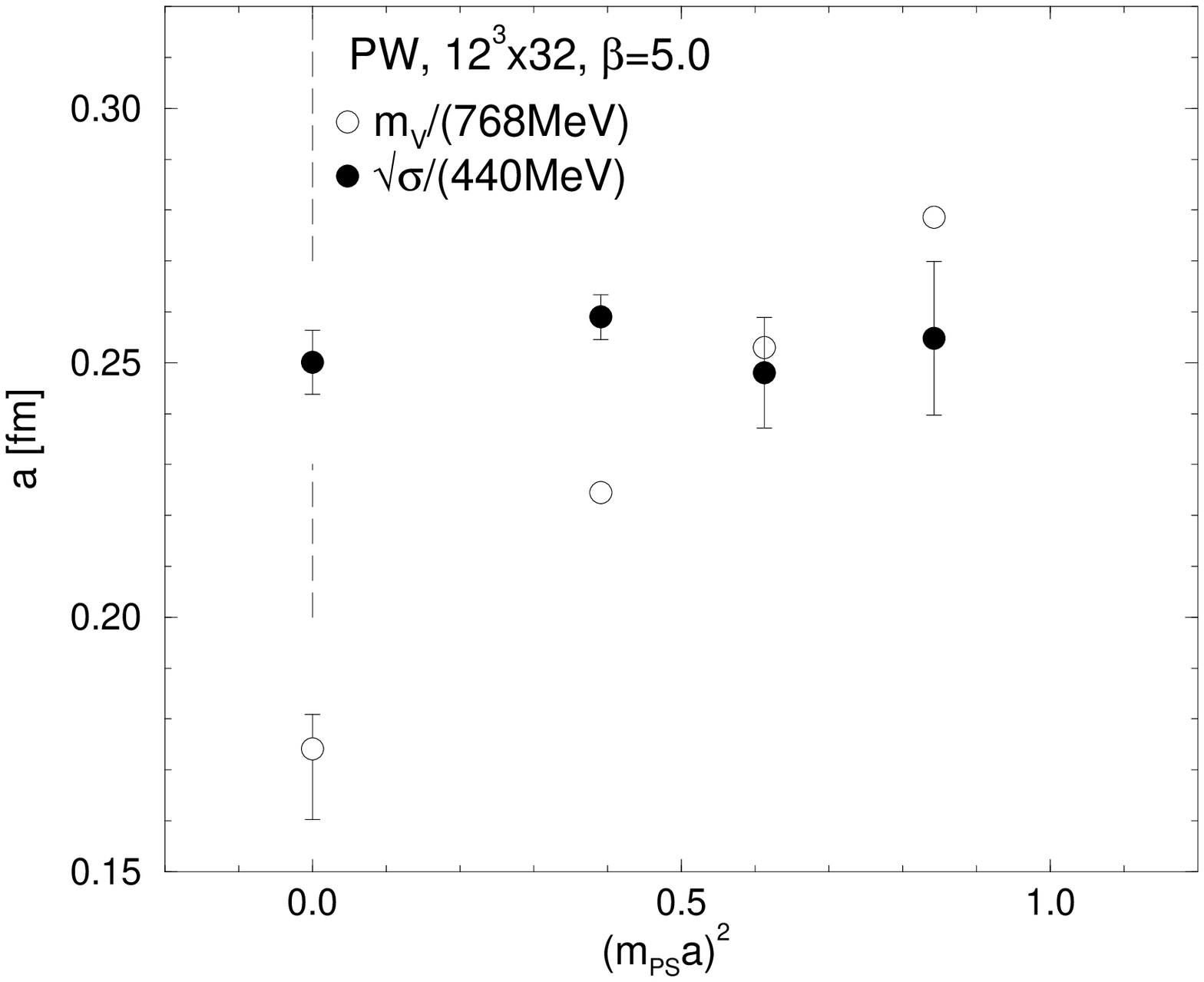}
\epsfxsize=8cm \epsfbox{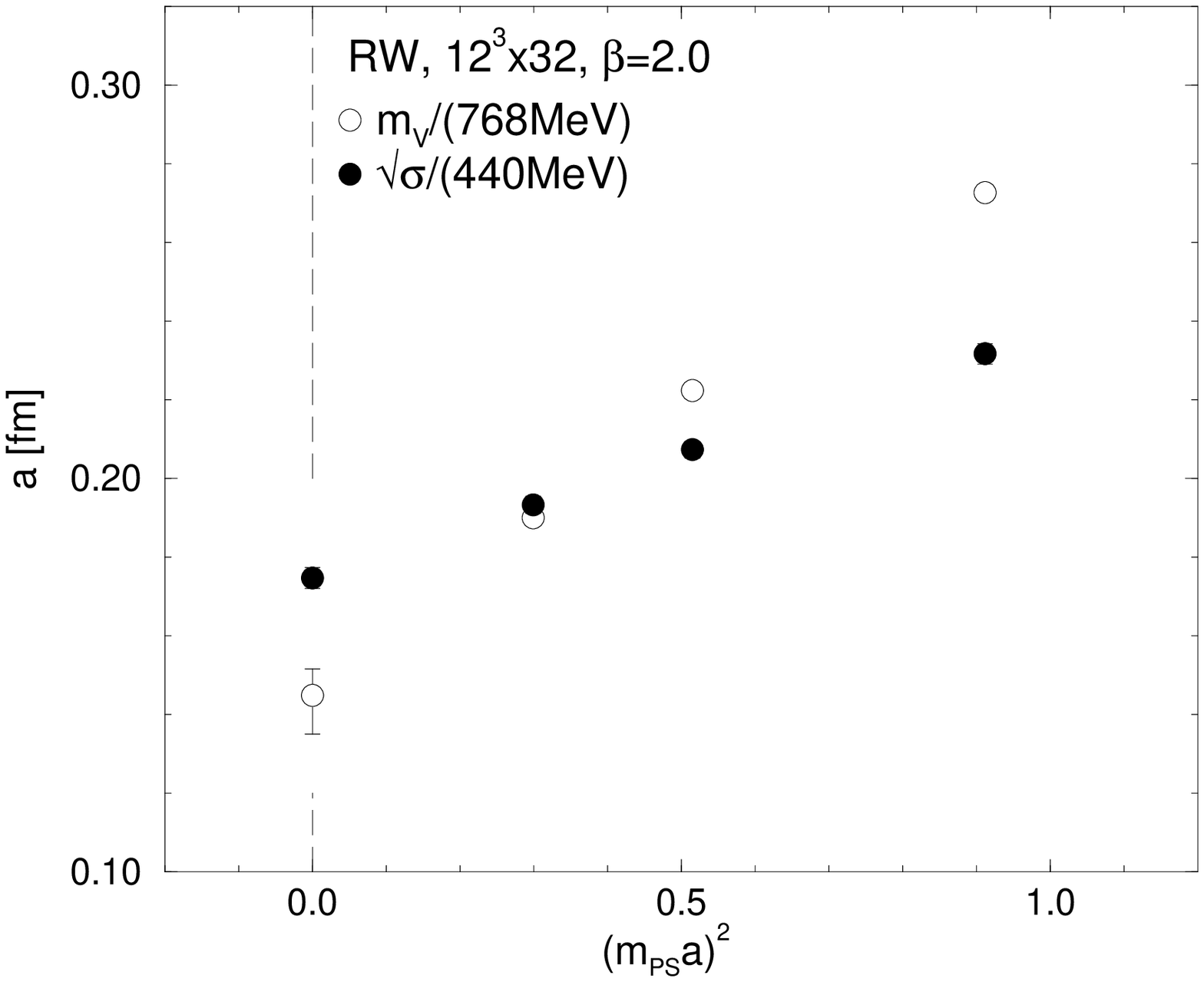}
}
\vspace{2mm}
\centerline{
\epsfxsize=8cm \epsfbox{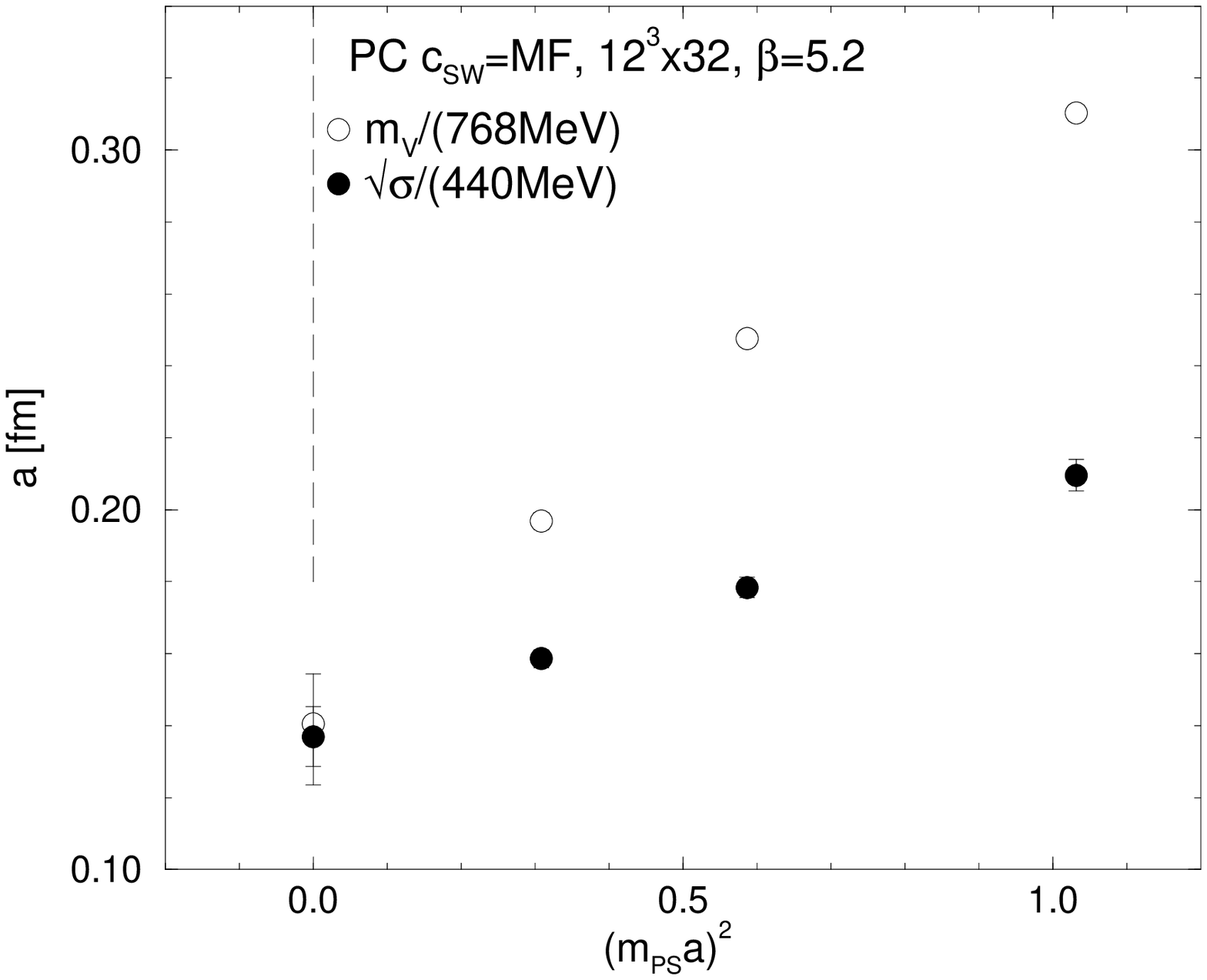}
\epsfxsize=8cm \epsfbox{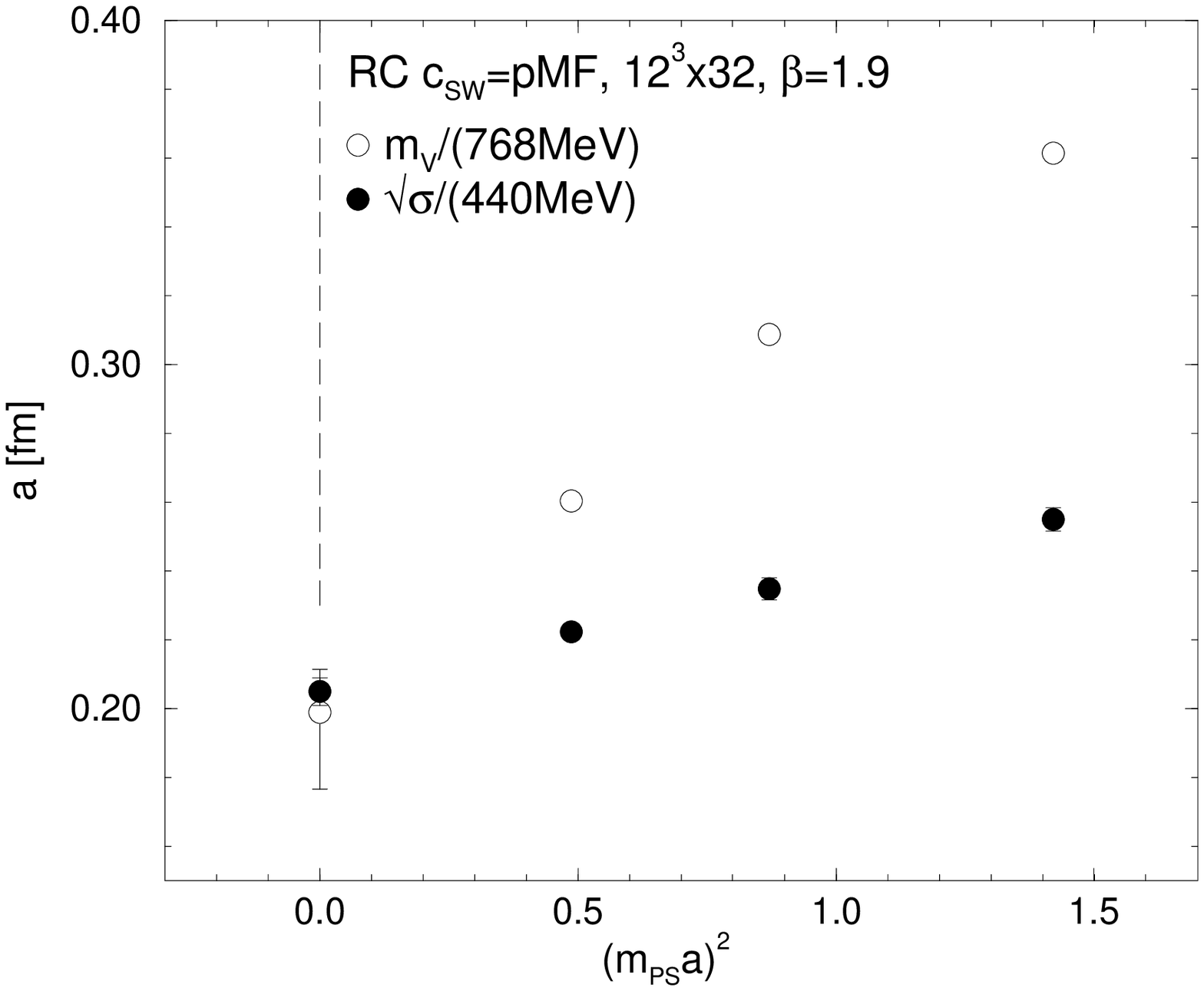}
}
\vspace{1mm}
\caption{Lattice spacing in physical units as calculated from 
$m_{\rm V}a/768$~MeV and $\sqrt{\sigma}a/440$~MeV as function of 
$(m_{\rm PS}a)^2$. Values in the chiral limit are also shown.}
\label{fig:a.all}
\vspace{0mm}
\end{figure*}

In Fig.~\ref{fig:a.all} we plot $m_{\rm V}a / 768{\rm MeV}$ and
$\sqrt{\sigma}a / 440{\rm MeV}$ as a function of $(m_{\rm PS}a)^2$ for the four
action combinations with a similar lattice spacing $a^{-1}_\rho\approx
1$--$1.3$~GeV determined from the vector meson mass.  A distinctive difference
between the results for the Wilson and the clover quark action is clear;
while results for $m_{\rm V}$ and $\sqrt{\sigma}$ cross each other at heavy
quark masses where $m_{\rm PS}/m_{\rm V}\approx 0.75$--$0.8$ for the \PW\ and
\RW\ cases, leading to a mismatch of $a_{\rho}$ and $a_{\sigma}$ in the
chiral limit, the two sets of physical scales converge well toward the
chiral limit for the \PC\ and \RC\ cases.

We expect the large discrepancy for the Wilson quark action
to disappear closer to the continuum limit. 
This is supported by the results obtained at $\beta=5.5$ with
$a^{-1} \approx 2$~GeV in Ref.\cite{SCRI-PW}.
Our results show that the clover term helps to improve
the consistency between $a_{\rho}$ and $a_{\sigma}$
already at $a^{-1} \approx 1$~GeV.

%////////////////////////////////////////////////////////////////////////////

\section{Approaching the chiral limit}
\label{sec:TXL}

The analyses presented so far show that the \RC\ action has 
the best scaling behavior for hadron masses and static quark potential 
among the four action combinations we have examined.  We then take this 
action and attempt to lower the quark mass as much as possible. 

Two runs are made at $\beta=1.9$, one on a $12^3\!\times\!32$ lattice down to 
$m_{\rm PS}/m_{\rm V}\approx 0.5$, and the other on a $16^3\!\times\!32$
lattice down to $m_{\rm PS}/m_{\rm V}\approx 0.4$. We discuss results 
from these runs below.

\subsection{hadrons with small quark masses}
\label{sec:small_mass}

\begin{figure*}[p]
\vspace{-1mm}
\centerline{
\epsfxsize=10cm \epsfbox{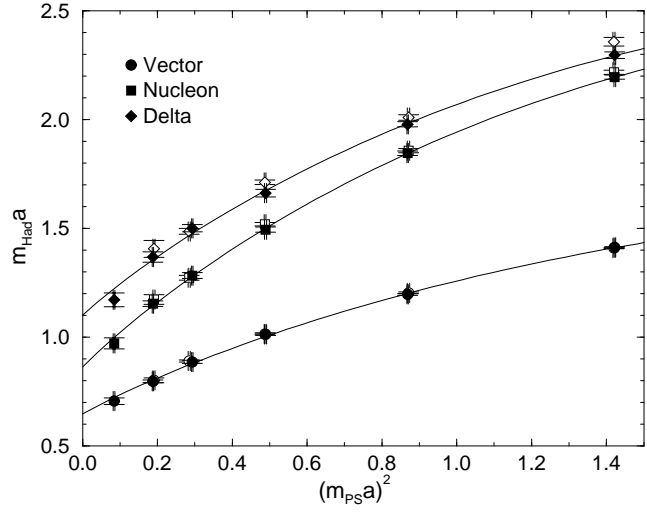}
}
\vspace{-40mm}
\caption{Chiral extrapolation of hadron masses as function of 
$(m_{\rm PS}a)^2$ for the \RCpmf\ action at $\beta=1.9$. Open symbols are 
results obtained on the $12^3\!\times\!32$ lattice whereas filled symbols
are from the $16^3\!\times\!32$ lattice. Lines are fits to the results for
the larger volume.}
\label{fig:chiralRC19}
\vspace{-4mm}
\end{figure*}

In Fig.~\ref{fig:chiralRC19} we plot the results of hadron masses as
functions of $(m_{\rm PS}a)^2$.  The existence of a curvature is observed, 
necessitating a cubic ansatz for extrapolation to the chiral limit. 
The lattice spacing determined from $m_\rho=768$~MeV equals 
$a_\rho=0.20(2)$~fm using mass results from the larger lattice.  
Hence the spatial size equals 2.4~fm ($12^3\!\times\!32$)
and 3.2~fm ($16^3\!\times\!32$) for the two lattice sizes employed. 

Finite-size effects are an important issue for precision determinations
of the hadron mass spectrum.  Our results in Fig.~\ref{fig:chiralRC19} do
not show clear signs of such effects down to the second lightest mass, 
which corresponds to $m_{\rm PS}/m_{\rm V}\approx 0.5$.  
We feel, however, that it is premature to draw conclusions with the 
present low statistics of approximately 1000 trajectories.

\begin{figure*}[p]
\vspace{-1mm}
\centerline{
\epsfxsize=10cm \epsfbox{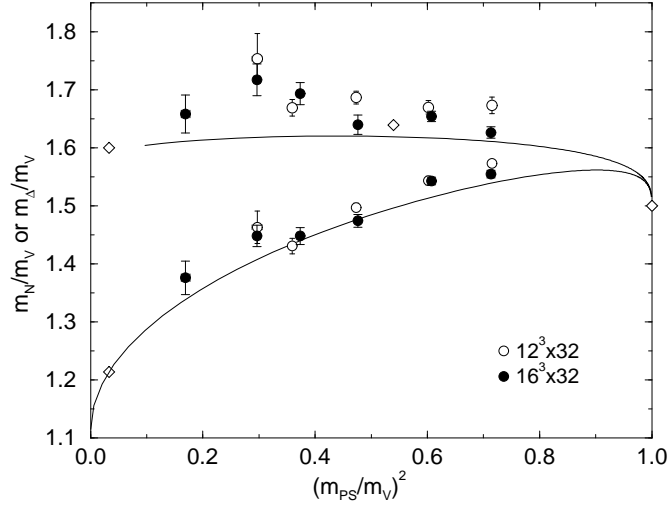}
}
\vspace{-40mm}
\caption{$m_{\rm N}/m_{\rm V}$ and $m_{\Delta}/m_{\rm V}$ as function of
$(m_{\rm PS}/m_{\rm V})^2$ for the two runs with
the \RCpmf\ action at $\beta=1.9$.} 
\label{fig:apeRC19}
\vspace{-4mm}
\end{figure*}

The results for mass ratios are plotted  in Fig.~\ref{fig:apeRC19}.  
While errors are large, and may even be underestimated because of the 
shortness of the runs, we find it encouraging that the ratios exhibit 
a trend of following the phenomenological curve toward the experimental 
points as the quark mass decreases. 
If we use the chiral extrapolation described above for the results on the
$16^3\!\times\!32$ lattice,  we obtain 
$m_{\rm N}/m_{\rm V} = 1.342(25)$  and $m_{\rm \Delta}/m_{\rm V} =
1.700(33)$ at the physical ratio $m_{\rm PS}/m_{\rm V} = 0.1757$, 
which are less than 10\% off the experimentally observed ratios of 1.223 
and 1.603, respectively,
despite the coarse lattice spacing of $a\approx 0.2$~fm.

\subsection{static potential at large distances}

We have mentioned in Sec.~\ref{sec:potential} that our results for the
static potential do not show signs of flattening, indicative of string 
breaking up to the distance of $r\approx 1$~fm.  Similar results have been 
reported by other groups \cite{Lat98.Kuti}.
A possible reason for these results is that potential data do not 
extend to large enough distances where string breaking becomes 
energetically favorable. Another related possibility is that the 
dynamical quark 
masses, which in most cases correspond to $m_{\rm PS}/m_{\rm V}=0.7$--$0.9$, 
are too heavy. With our runs on the $16^3\!\times\!32$ lattice 
we can examine these points up to the distance of $r\approx 2$~fm and 
for quark masses down to  $m_{\rm PS}/m_{\rm V} \approx 0.4$. 

\begin{figure*}[t]
\vspace{-1mm}
\centerline{
\epsfxsize=10cm \epsfbox{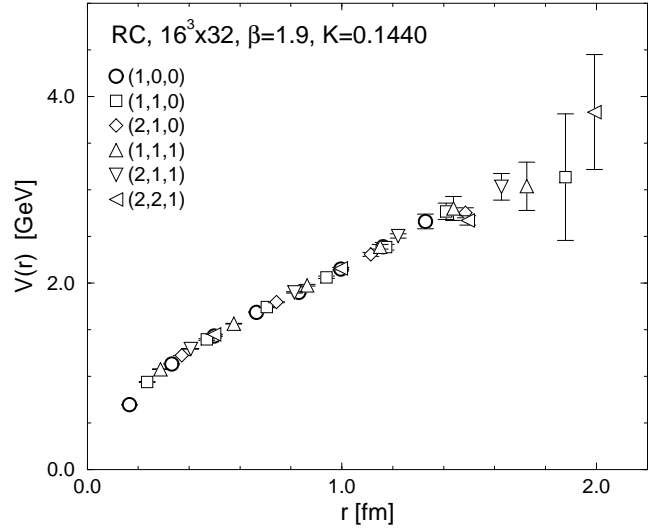}
}
\vspace{-1mm}
\caption{Static quark potential on the $16^3\!\times\!32$ lattice
at the lightest sea quark mass $m_{\rm PS}/m_{\rm V} \approx
0.4$. The scale is set by $a_{\rho}$ in the chiral limit.
}
\label{fig:VvsR.16x32}
\end{figure*}

In Fig.~\ref{fig:VvsR.16x32}
we plot our potential data obtained 
on the $16^3\!\times\!32$ lattice 
at the lightest sea quark mass corresponding to 
$m_{\rm PS}/m_{\rm V} \approx 0.4$.
We find that the potential increases
linearly up to $r \approx 2$~fm,
without any clear signal of flattening. 
The situation is similar for our data
at heavier sea quark masses.

\begin{figure*}[ht]
\vspace{3mm}
\centerline{
\epsfxsize=10cm \epsfbox{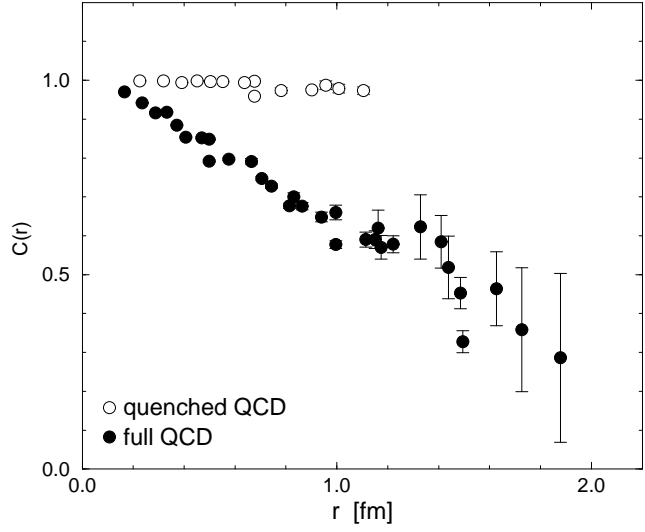}
}
\vspace{0mm}
\caption{Overlap function $C(R)$ for full and quenched QCD as a function of
$r$. Filled symbols are the data in full QCD
on the $16^3\!\times\!32$ lattice with the \RCpmf\ action  
at $\beta=1.9$ and $K=0.1440$.
Open symbols represent data in quenched QCD
on a $9^3\!\times\!18$ lattice with the RG improved 
gauge action at $\beta = 2.1508$ ($a^{-1} \approx 1$~GeV).}
\label{fig:CvsR}
\vspace{-4mm}
\end{figure*}

An interesting and crucial question here 
is whether the Wilson loop operator has 
sufficient overlap with the ground state at large $r$
so that the potential in that state is reliably measured
there \cite{lat97.pot}.
In Fig.~\ref{fig:CvsR} we compare results for the overlap function $C(r)$
for the full QCD run at $m_{\rm PS}/m_{\rm V} \approx 0.4$ 
with that obtained in a quenched run with the \R\ gauge action on 
a $9^3\!\times\!18$
lattice at $\beta=2.1508$ ($a^{-1} \approx 1$~GeV) \cite{kaneko.Tcs}.
For the quenched run the overlap $C(r)$ of the smeared Wilson loop operator 
with the ground string state is effectively 100~\% 
at all distances.
For full QCD, on the other hand, $C(r)$ significantly 
decreases as $r$ increases.  Such a behavior of $C(r)$ is observed 
in all of our data including those taken with action choices other than
\RC .  
These results may be taken as a tantalizing hint that the Wilson loop 
operator develops mixings with states other than a single string, 
possibly a pair of static-light mesons in full QCD.
We leave further investigations of this interesting 
question for future studies. 

\subsection{computer time}

An important practical information in full QCD is the computer time needed 
for the approach to the chiral limit. 
In Table \ref{tab:CPUtime} we assemble the relevant numbers for our
runs on the $16^3\!\times\!32$ lattice. 
These runs have been performed on a partition of 256 nodes, which is 1/8 of
the CP-PACS computer. For a partition of this size, our full QCD program,
written in FORTRAN with the matrix multiplication in the quark solver
hand-optimized in the assembly language, sustains about 37\% of the peak
speed of 75 GFLOPS.
Adding the CPU time per trajectory of Table \ref{tab:CPUtime}, we find 
that accumulating 5000 trajectories for each of the 6 hopping parameters 
for this lattice size would take about 160 days with the full use of 
the CP-PACS computer. Carrying out such a simulation is certainly feasible.
For larger lattice sizes 
such as $24^3\!\times\!48$, however, we would have to stop at 
$m_{\rm PS}/m_{\rm V}\approx 0.5$ since the run at 
$m_{\rm PS}/m_{\rm V}\approx 0.4$ alone increases the computer time 
by a factor two.      
Let us add that the CPU time for a 
unit of HMC trajectory increases roughly proportional to $(1/K-1/K_c)^{-1.6}$ 
for the 4 smallest quark masses.

%////////////////////////////////////////////////////////////////////////////

\section{Conclusions}
\label{sec:conclusion}

In this paper we have presented a detailed investigation of the effect of
improving the gauge and the quark action in full QCD.
We have found that the consequence of improving either of the actions is
different depending on the observable examined.

For the light hadron spectrum the clover quark action with a
mean-field improved coefficient drastically improves the scaling of hadron
mass ratios.  Improving the gauge action, on the other hand, has almost
no influence in this aspect.  The SW-clover action also has the good 
property that the physical scale determined
from the vector meson mass and the string tension in the chiral limit of
the sea quark are consistent already at scales
$a^{-1} \approx 1$~GeV, which is not the case with the Wilson quark action. 

We have also confirmed that the use of improved gauge actions leads to
a significant decrease of the breaking of rotational symmetry of
the static quark potential.

Finally, we have made an exploratory simulation toward the chiral limit
employing a renormalization group improved gauge and clover improved quark
actions.

The results obtained in the present study suggest that a significant step
toward a systematic full QCD simulation can be made with the present
computing power using improved gauge and quark actions
at relatively coarse lattice spacings of $a^{-1}\approx 1$--2~GeV.

\acknowledgements

This work was supported in part by 
the Grants-in-Aid of the Ministry of Education 
(Nos.\ 08NP0101, 08640349, 08640350, 08640404, 08740189, 08740221,
09304029, 10640246 and 10640248).  Two of us (GB, RB) were supported by 
the Japan Society for the Promotion of Science.

%%%%%%%%%%%%%% tables %%%%%%%%%%%%%%%%

\begin{table}[tb]
\setlength{\tabcolsep}{0.25pc}
\caption{Overview of the simulations on the $12^3\!\times\!32$ lattice for the
action comparison.}
\label{tab:run-overview}
\vspace{2mm}
\begin{center} 
\begin{tabular}{lllllll}
action & $\beta$ & $c_{\rm SW}$ & $m_{\rm PS}/m_{\rm V}$ & $K_c$ & 
${a_{\rho}}$ [fm] & ${a_{\sigma}}$ [fm] \\
\hline
\PW\    & 4.8  & --           & 0.83,0.77,0.70 & 0.19286(14) & 0.197(2)        
    & --         \\
\PW\    & 5.0  & --           & 0.85,0.79,0.71 & 0.18291(7)  & 0.174$^{+8}_{-22}$   
    & 0.2501(62) \\
\RW\    & 1.9  & --           & 0.90,0.80,0.69 & 0.17398(7)  & 0.162$^{+11}_{-15}$  
    & --         \\
\RW\    & 2.0  & --           & 0.90,0.83,0.74 & 0.16726(8)  & 0.144$^{+7}_{-13}$   
    & 0.1747(27) \\
\PCtr\  & 5.0  & 1.0          & 0.83,0.79,0.71 & 0.16631(18) & 0.2157(4)       
    & --         \\
\PCmf\  & 5.0  & 1.805--1.855 & 0.81,0.76,0.71 & 0.14927(28) & 0.238(1)        
    & 0.241(12)  \\
\PCmf\  & 5.2  & 1.64--1.69   & 0.84,0.79,0.72 & 0.14298(6)  & 0.141$^{+15}_{-24}$  
    & 0.1370(83) \\
\PCmf\  & 5.25 & 1.61--1.637  & 0.84,0.76      & 0.14252(4) & 0.133(3)        
    & 0.1161(89) \\
\RCpmf\ & 1.9  & 1.55         & 0.85,0.78,0.69 & 0.14446(6) & 0.199$^{+14}_{-27}$  
    & 0.2050(40) \\
\RCtr\  & 2.0  & 1.0          & 0.88,0.83,0.71 & 0.15045(10) & 0.160$^{+10}_{-18}$  
    & 0.1638(42) \\
\RCmf\  & 2.0  & 1.515--1.54  & 0.90,0.86,0.79,0.70 & 0.14083(4) & 0.146(3)    
    & 0.152(3)  \\
\RCpmf\ & 2.0  & 1.505        & 0.91,0.79,0.71 & 0.14058(7) & 0.146$^{+35}_{-22}$  
    & --         \\
\end{tabular}
\end{center}  
\end{table}

\begin{table}[tb]
\setlength{\tabcolsep}{0.23pc}
\caption{Overview of the simulations exploring the chiral limit of full QCD.}
\label{tab:TXL-overview}
\vspace{2mm}
\begin{center} 
\begin{tabular}{lllllll}
size & $\beta$ & $c_{\rm SW}$ & $m_{\rm PS}/m_{\rm V}$ & $K_c$ & ${a_{\rho}}$
[fm] 
& ${a_{\sigma}}$ [fm] \\
\hline
$12^3\!\times\!32$  & 1.9 & 1.55 & 0.85,0.78,0.69,0.60,0.54  & 0.144432(18)   
   & 0.171(3) & --           \\
$16^3\!\times\!32$  & 1.9 & 1.55 & 0.84,0.78,0.69,0.61,0.54,0.41 & 0.144434(10)
   & 0.166(2)   & 0.1817(28)   \\
\end{tabular}
\end{center}  
\end{table}

\begin{table}[htb]
\setlength{\tabcolsep}{0.3pc}
\caption{CPU time per HMC trajectory 
for the run at $\beta=1.9$ on the $16^3\!\times\!32$ lattice
carried out on CP-PACS with 256 nodes (75 GFLOPS peak).}
\label{tab:CPUtime}
\vspace{2mm}
\begin{center} 
\begin{tabular}{llllllrr}
$K$ & $(1/K-1/K_c)/2$ & $m_{\rm PS}/m_{\rm V}$  & 
$\Delta\tau$ & accept. & stop & $N_{\rm inv}$ & CPU-time \\
\hline
0.1370 & 0.1879(2)& 0.8446(15)& 0.0075 & 0.86 &$10^{-11}$&  30 &   6.4 min. \\
0.1400 & 0.1096(2)& 0.7793(19)& 0.0075 & 0.80 &$10^{-11}$&  46 &   8.2 min. \\
0.1420 & 0.0593(2)& 0.6899(33)& 0.00625& 0.77 &$10^{-11}$&  74 &  14.2 min. \\
0.1430 & 0.0347(2)& 0.6110(44)& 0.004  & 0.77 &$10^{-11}$& 116 &  32.3 min. \\
0.1435 & 0.0225(2)& 0.5445(50)& 0.0025 & 0.81 &$10^{-12}$& 181 &  77.6 min. \\
0.1440 & 0.0104(2)& 0.4115(96)& 0.0015 & 0.66 &$10^{-12}$& 344 & 230.4 min. \\
\end{tabular}
\end{center}  
\end{table}

\appendix

\section{Run Parameters}
\label{appendix:param}

In this appendix we assemble information about our 
runs. An overview of the runs has been given in
Table~\ref{tab:run-overview}. 
For the inversion of the quark matrix either the MR algorithm (M) or the
BiCGStab algorithm (B) is used with the stopping condition $r_1 \leq$~stop 
defined through Eq.(\ref{eq:stopp}). 
During the HMC update $D^{\dagger}D$ has to be inverted. We do this in two
steps, first inverting $D^{\dagger}$ and then $D$. In the tables we quote
the number of iterations $N_{\rm inv}$ needed for the first inversion
$D^{\dagger}$. 
Finally we also quote the statistics, giving the number of configurations
for spectrum and potential measurements separately.
Configurations for the hadron spectrum are separated by 5 HMC
trajectories, whereas for the potential the separation is either 5 
or 10 trajectories.
Unless stated otherwise the lattice size is $12^3\!\times\!32$.

\begin{table}[htb]
\caption{Simulation parameters for the \PW\ and \RW\ action combination.} 
\label{tab:paramPRW}
\vspace{2mm}
\begin{center} 
\begin{tabular}{lllllllrrr}
 action & $\beta$ & $K$ & $\Delta\tau$ & accept.& inverter  & stop & \
 $N_{\rm inv}$  & \#conf & \#conf$\times$sep \\
        &         &     &              &     &           &        & \
                & spect. & pot. \\
\hline
\PW\ & 4.8   & 0.1846  & 0.01   & 0.78 & M & $10^{-10}$ & 100  & 222 &  -- \\  
     &       & 0.1874  & 0.005  & 0.88 & M & $10^{-10}$ & 150  & 200 &  -- \\ 
     &       & 0.1891  & 0.005  & 0.83 & M & $10^{-10}$ & 199  & 200 &  -- \\ 
\cline{2-10}                       
     & 5.0   & 0.1779  & 0.01   & 0.79 & M & $10^{-10}$ & 101  & 300 & $ 89 \times 5$ \\ 
     &       & 0.1798  & 0.005  & 0.94 & M & $10^{-10}$ & 147  & 301 & $100 \times 5$ \\
     &       & 0.1811  & 0.005  & 0.88 & M & $10^{-10}$ & 212  & 301 & $100 \times 5$ \\
\hline                       
\RW\ & 1.9   & 0.1632  & 0.0125 & 0.82 & M & $10^{-10}$ &  73  & 200 &  -- \\
     &       & 0.1688  & 0.01   & 0.78 & M & $10^{-10}$ & 136  & 200 & $100 \times 5$ \\
     &       & 0.1713  & 0.008  & 0.71 & M & $10^{-10}$ & 234  & 200 &  -- \\
\cline{2-10}                       
     & 2.0   & 0.1583  & 0.0125 & 0.79 & M & $10^{-10}$ &  77  & 300 & $100 \times 5$ \\
     &       & 0.1623  & 0.01   & 0.84 & M & $10^{-10}$ & 128  & 300 & $100 \times 5$ \\
     &       & 0.1644  & 0.008  & 0.82 & M & $10^{-10}$ & 212  & 305 & $ 96 \times 5$ \\
\end{tabular}
\end{center}  
\end{table}

\begin{table}[htb]
\setlength{\tabcolsep}{0.3pc}
\caption{Simulation parameters for the \PC\ action combination. }
\label{tab:paramPC}
\vspace{2mm}
\begin{center} 
\begin{tabular}{lllllllrrr}
 $\beta$ & $K$ & $c_{\rm SW}$ & $\Delta\tau$ & accept. & inverter & stop & \
 $N_{\rm inv}$ & \#conf  & \#conf$\times$sep \\
         &     &        &              &      &          &        & \
               & spect.   & pot. \\
\hline
 5.0  & 0.1590 & 1.0   & 0.01   & 0.82 & B & $10^{-10}$ & 37  & 100 & -- \\
      & 0.1610 & 1.0   & 0.008  & 0.83 & B & $10^{-10}$ & 44  & 100 & -- \\
      & 0.1630 & 1.0   & 0.00625& 0.80 & B & $10^{-10}$ & 67  & 101 & -- \\
\hline       
 5.0  & 0.1415 & 1.855 & 0.01   & 0.73 & B & $10^{-10}$ & 30  & 200 & $100\times 10$ \\
      & 0.1441 & 1.825 & 0.008  & 0.75 & B & $10^{-10}$ & 42  & 200 & $100\times 10$ \\
      & 0.1455 & 1.805 & 0.00625& 0.77 & B & $10^{-10}$ & 55  & 200 & $100\times 10$ \\
\hline          
 5.2  & 0.1390 & 1.69  & 0.01   & 0.81 & M & $10^{-10}$ &  72 & 248 & $104\times 5$ \\ 
      & 0.1410 & 1.655 & 0.008  & 0.83 & M & $10^{-10}$ & 117 & 232 & $100\times 5$ \\
      & 0.1420 & 1.64  & 0.008  & 0.73 & M & $10^{-10}$ & 203 & 200 & $100\times 5$ \\
\hline          
 5.25 & 0.1390 & 1.637 & 0.008  & 0.88 & M & $10^{-10}$ &  88 & 198 & $ 69\times 5$ \\
      & 0.1410 & 1.61  & 0.00667& 0.84 & M & $10^{-10}$ & 183 & 194 & $101\times 5$ \\
\end{tabular}
\end{center}  
\end{table}

\begin{table}[htb]
\setlength{\tabcolsep}{0.3pc}
\caption{Simulation parameters for the \RC\ action combination. 
The run marked with (*) is on the $16^3\!\times\!32$ lattice.}
\label{tab:paramRC}
\vspace{2mm}
\begin{center} 
\begin{tabular}{lllllllrrr}
 $\beta$ & $K$ & $c_{\rm SW}$ & $\Delta\tau$ & accept. & inv. \
& stop & $N_{\rm inv}$ & \#conf  & \#conf$\times$sep \\
         &     &        &              &      &      \ 
&        &               & spect.  & pot. \\
\hline
$1.9^*$& 0.1370 & 1.55 & 0.0075 & 0.86 & B &$10^{-11}$&  30  & 203  & --  \\  
       & 0.1400 & 1.55 & 0.0075 & 0.80 & B &$10^{-11}$&  46  & 198  & --  \\
       & 0.1420 & 1.55 & 0.00625& 0.77 & B &$10^{-11}$&  74  & 202  &  $92\times 10$ \\
       & 0.1430 & 1.55 & 0.004  & 0.77 & B &$10^{-11}$& 116  & 212  & $102\times 10$ \\
       & 0.1435 & 1.55 & 0.0025 & 0.81 & B &$10^{-12}$& 181  & 263  & --  \\
       & 0.1440 & 1.55 & 0.0015 & 0.66 & B &$10^{-12}$& 344  &  79  &  $79\times 10$ \\
\hline          
 1.9   & 0.1370 & 1.55 & 0.01   & 0.82 & B &$10^{-10}$&  28  & 267  & $127\times 10$ \\
       & 0.1400 & 1.55 & 0.01   & 0.78 & B &$10^{-10}$&  41  & 214  & $104\times 10$ \\
       & 0.1420 & 1.55 & 0.008  & 0.72 & B &$10^{-10}$&  66  & 324  & $148\times 10$ \\
       & 0.1430 & 1.55 & 0.005  & 0.77 & B &$10^{-10}$& 102  & 302  & --  \\
       & 0.1435 & 1.55 & 0.00333& 0.79 & B &$10^{-11}$& 159  & 170  & --  \\
\hline             
 2.0   & 0.1420 & 1.0  & 0.01   & 0.87 & B &$10^{-10}$&  29  & 100  & $50\times 10$ \\
       & 0.1450 & 1.0  & 0.008  & 0.91 & B &$10^{-10}$&  42  & 100  & $50\times 10$ \\
       & 0.1480 & 1.0  & 0.00625& 0.86 & B &$10^{-10}$&  81  & 100  & $50\times 10$ \\
\hline             
 2.0   & 0.1300 & 1.505 & 0.01  & 0.90 & B &$10^{-10}$&  21  & 100  & -- \\  
       & 0.1370 & 1.505 & 0.008 & 0.86 & B &$10^{-10}$&  47  &  90  & -- \\
       & 0.1388 & 1.505 & 0.008 & 0.78 & B &$10^{-10}$&  79  &  90  & -- \\
\hline             
 2.0   & 0.1300 & 1.54 & 0.008  & 0.93 & M  &$10^{-10}$& 42   & 201 & $100\times 5$ \\
       & 0.1340 & 1.529& 0.008  & 0.90 & M  &$10^{-10}$& 62   & 200 & $100\times 10$ \\
       & 0.1370 & 1.52 & 0.008  & 0.87 & M/B&$10^{-10}$&102/50& 200 & $102\times 5$ \\
       & 0.1388 & 1.515& 0.00625& 0.84 & M/B&$10^{-10}$&181/84& 200 & $105\times 5$ \\
\end{tabular}
\end{center}  
\end{table}

\section{Hadron Masses}
\label{appendix:mass}

In this appendix we assemble the results of our hadron mass measurements.
We quote numbers for 
pseudo scalar and vector mesons, nucleons and $\Delta$ baryons together
with mass ratios against vector mesons. Additionally we quote numbers
for the bare quark mass based on the axial Ward identity
defined by
\begin{equation}
m_qa = -m_{\rm PS}a \lim_{t\rightarrow\infty} 
\frac{\sum_{\vec{x}} \langle A_4(\vec{x},t)P\rangle} 
     {\sum_{\vec{x}} \langle   P(\vec{x},t)P\rangle},
\end{equation}
where $A_4$ is the local axial current and $P$ is the pseudo scalar
density. Masses are extracted with an uncorrelated fit to the propagator and
the errors are determined with the jackknife method with bin size
5.

\begin{table}[htb]
\caption{\PW\ action combination: AWI quark mass and meson masses.}
\label{tab:mesonPW}
\vspace{2mm}
\begin{center} 
\begin{tabular}{llllll}
 $\beta$ & $K$ & $m_qa$ & $m_{\rm PS}a$ & $m_{\rm V}a$ & 
 $m_{\rm PS}/m_{\rm V}$ \\
\hline
 4.8   & 0.1846  & 0.13400(68) & 0.9350(9)  & 1.1276(18)  & 0.8291(12)  \\
       & 0.1874  & 0.09269(80) & 0.7918(13) & 1.0263(25)  & 0.7715(17)  \\ 
       & 0.1891  & 0.06523(70) & 0.6716(16) & 0.9559(45)  & 0.7026(32)  \\ 
\hline
 5.0   & 0.1779  & 0.13464(91) & 0.9182(10) & 1.0859(17)  & 0.8456(12)  \\ 
       & 0.1798  & 0.09652(88) & 0.7829(14) & 0.9863(23)  & 0.7938(18)  \\ 
       & 0.1811  & 0.0610(12)  & 0.6254(32) & 0.8753(38)  & 0.7145(42)  \\ 
\end{tabular}
\end{center}  
\end{table}

\begin{table}[htb]
\caption{\RW\ action combination: AWI quark mass and meson masses.}
\label{tab:mesonRW}
\vspace{2mm}
\begin{center} 
\begin{tabular}{llllll}
 $\beta$ & $K$ & $m_qa$ & $m_{\rm PS}a$ & $m_{\rm V}a$ & 
 $m_{\rm PS}/m_{\rm V}$ \\
\hline
 1.9   & 0.1632  & 0.1972(15)  & 1.0557(11) & 1.1743(16) & 0.8990(9)   \\
       & 0.1688  & 0.0977(13)  & 0.7525(19) & 0.9377(35) & 0.8025(26)  \\
       & 0.1713  & 0.05281(84) & 0.5469(21) & 0.7935(52) & 0.6892(43)  \\
\hline     
 2.0   & 0.1583  & 0.1761(11)  & 0.9551(12) & 1.0631(17) & 0.8984(90)  \\
       & 0.1623  & 0.10021(88) & 0.7177(14) & 0.8671(27) & 0.8277(20)  \\
       & 0.1644  & 0.06010(61) & 0.5475(16) & 0.7406(27) & 0.7394(26)  \\
\end{tabular}
\end{center}  
\end{table}

\begin{table}[htb]
\caption{\PC\ action combination: AWI quark mass and meson masses.}
\label{tab:mesonPC}
\vspace{2mm}
\begin{center} 
\begin{tabular}{llllll}
 $\beta$ & $K$ & $m_qa$ & $m_{\rm PS}a$ & $m_{\rm V}a$ & 
 $m_{\rm PS}/m_{\rm V}$ \\
\hline
$5.0_{\rm tree}$ &0.1590& 0.2029(17) & 1.1105(10) & 1.3452(36) & 0.8256(21)\\ 
                 &0.1610& 0.1509(17) & 0.9641(28) & 1.2193(69) & 0.7907(38)\\  
                 &0.1630& 0.0956(20) & 0.7740(22) & 1.0865(81) & 0.7124(60)\\  
           \cline{2-6}   
$5.0_{\rm MF}$ & 0.1415 & 0.2211(17) & 1.1970(18) & 1.4769(44) & 0.8104(26)\\  
               & 0.1441 & 0.1574(15) & 0.9961(19) & 1.3156(65) & 0.7571(36)\\  
               & 0.1455 & 0.1176(15) & 0.8588(42) & 1.2024(99) & 0.7143(44)\\  
\hline   
$5.2$  & 0.1390 & 0.1855(24) & 1.0161(27) & 1.2100(48) & 0.8398(20)  \\  
       & 0.1410 & 0.1160(17) & 0.7662(43) & 0.9654(72) & 0.7937(30)  \\  
       & 0.1420 & 0.0646(24) & 0.5553(55) & 0.7674(93) & 0.7236(76)  \\  
\hline   
$5.25$ & 0.1390 & 0.1435(19) & 0.8479(30) & 1.0155(42) & 0.8350(26)  \\  
       & 0.1410 & 0.0731(17) & 0.5532(42) & 0.7296(91) & 0.7581(57)  \\  
\end{tabular}
\end{center}  
\end{table}

\begin{table}[htb]
\caption{\RC\ action combination: AWI quark mass and meson masses.
The run marked with (*) is on the $16^3\!\times\!32$ lattice.}
\label{tab:mesonRC}
\vspace{2mm}
\begin{center} 
\begin{tabular}{llllll}
 $\beta$ & $K$ & $m_qa$ & $m_{\rm PS}a$ & $m_{\rm V}a$ & 
 $m_{\rm PS}/m_{\rm V}$ \\
\hline
$1.9^{*}$& 0.1370 & 0.2428(10)  & 1.1926(11) & 1.4121(31) & 0.8446(15)  \\ 
         & 0.1400 & 0.1517(10)  & 0.9321(11) & 1.1961(36) & 0.7793(19)  \\ 
         & 0.1420 & 0.08834(88) & 0.6992(19) & 1.0134(60) & 0.6899(33)  \\ 
         & 0.1430 & 0.05530(62) & 0.5414(18) & 0.8861(71) & 0.6110(44)  \\ 
         & 0.1435 & 0.03484(75) & 0.4338(20) & 0.7967(68) & 0.5445(50)  \\ 
         & 0.1440 & 0.0156(15)  & 0.2906(41) & 0.706(15)  & 0.4115(96)  \\ 
\hline
1.9      & 0.1370 & 0.2440(13)  & 1.1918(12) & 1.4091(28) & 0.8458(17)  \\ 
         & 0.1400 & 0.1547(10)  & 0.9334(17) & 1.2033(39) & 0.7757(18)  \\ 
         & 0.1420 & 0.08975(96) & 0.6983(18) & 1.0149(45) & 0.6880(31)  \\
         & 0.1430 & 0.05278(77) & 0.5337(24) & 0.8902(53) & 0.5995(38)  \\ 
         & 0.1435 & 0.0374(17)  & 0.4368(30) & 0.802(10)  & 0.5448(82)  \\
\hline   
$2.0_{\rm tree}$ &0.1420& 0.2303(14) & 1.0888(22) & 1.2403(33) & 0.8779(15)\\ 
                 &0.1450& 0.1519(13) & 0.8645(28) & 1.0415(44) & 0.8300(21)\\ 
                 &0.1480& 0.0713(16) & 0.5730(24) & 0.8064(79) & 0.7105(59)\\ 
             \cline{2-6}   
$2.0_{\rm pMF}$  &0.1300& 0.3313(18) & 1.3358(21) & 1.4682(33) & 0.9098(11)\\ 
                 &0.1370& 0.1305(10) & 0.7784(25) & 0.9801(47) & 0.7942(31)\\ 
                 &0.1388& 0.0665(13) & 0.5489(38) & 0.773(11)  & 0.7098(77)\\ 
             \cline{2-6}   
$2.0_{\rm MF}$ & 0.1300 & 0.3158(10) & 1.2971(11) & 1.4377(22) & 0.9022(11) \\ 
               & 0.1340 & 0.2079(10) & 1.0137(17) & 1.1759(27) & 0.8620(16) \\ 
               & 0.1370 & 0.1190(10) & 0.7435(17) & 0.9400(44) & 0.7910(32) \\ 
               & 0.1388 & 0.0671(10) & 0.5416(24) & 0.7741(71) & 0.6997(56) \\ 
\end{tabular}
\end{center}  
\end{table}

\begin{table}[htb]
\caption{\PW\ action combination: baryon masses.}
\label{tab:baryonPW}
\vspace{2mm}
\begin{center} 
\begin{tabular}{llllll}
 $\beta$ & $K$ & $m_{\rm N}a$ & $m_{\Delta}a$ & $m_{\rm N}/m_{\rm V}$ & 
 $m_\Delta/m_{\rm V}$ \\
\hline
 4.8   & 0.1846  & 2.009(12) & 2.074(15) & 1.782(11) & 1.839(13)  \\
       & 0.1874  & 1.817(18) & 1.912(23) & 1.771(18) & 1.863(23)  \\ 
       & 0.1891  & 1.647(20) & 1.848(32) & 1.723(22) & 1.933(36)  \\ 
\hline
 5.0   & 0.1779  & 1.894(12) & 1.976(17) & 1.744(11) & 1.819(16)  \\ 
       & 0.1798  & 1.668(15) & 1.775(13) & 1.691(14) & 1.799(12)  \\ 
       & 0.1811  & 1.437(17) & 1.559(18) & 1.642(20) & 1.781(19)  \\ 
\end{tabular}
\end{center}  
\end{table}

\begin{table}[htb]
\caption{\RW\ action combination: baryon masses.}
\label{tab:baryonRW}
\vspace{2mm}
\begin{center} 
\begin{tabular}{llllll}
 $\beta$ & $K$ & $m_{\rm N}a$ & $m_{\Delta}a$ & $m_{\rm N}/m_{\rm V}$ & 
 $m_\Delta/m_{\rm V}$ \\
\hline
 1.9   & 0.1632  & 1.997(14)  & 2.044(15)  & 1.700(12)  & 1.740(13)  \\
       & 0.1688  & 1.548(15)  & 1.650(21)  & 1.651(13)  & 1.760(20)  \\
       & 0.1713  & 1.2643(88) & 1.417(17)  & 1.593(12)  & 1.786(19)  \\
\hline     
 2.0   & 0.1583  & 1.7589(57) & 1.8150(77) & 1.6545(48) & 1.7073(62) \\
       & 0.1623  & 1.4214(77) & 1.5008(90) & 1.6392(80) & 1.7308(84) \\
       & 0.1644  & 1.1752(80) & 1.281(11)  & 1.587(10)  & 1.729(14)  \\
\end{tabular}
\end{center}  
\end{table}

\begin{table}[htb]
\caption{\PC\ action combination: baryon masses.}
\label{tab:baryonPC}
\vspace{2mm}
\begin{center} 
\begin{tabular}{llllll}
 $\beta$ & $K$ & $m_{\rm N}a$ & $m_{\Delta}a$ & $m_{\rm N}/m_{\rm V}$ & 
 $m_\Delta/m_{\rm V}$ \\
\hline
$5.0_{\rm tree}$ & 0.1590 & 2.203(25) & 2.358(30) & 1.638(20) & 1.753(23)  \\ 
                 & 0.1610 & 1.982(24) & 2.110(30) & 1.625(13) & 1.730(18)  \\  
                 & 0.1630 & 1.748(22) & 1.868(44) & 1.609(21) & 1.719(40)  \\  
                 \cline{2-6}  
$5.0_{\rm MF}$   & 0.1415 & 2.343(24) & 2.501(28) & 1.586(16) & 1.693(17)  \\  
                 & 0.1441 & 2.041(20) & 2.243(27) & 1.551(14) & 1.705(18)  \\  
                 & 0.1455 & 1.851(21) & 1.994(31) & 1.539(15) & 1.659(24)  \\  
\hline   
5.2        & 0.1390 & 1.864(13) & 1.980(16) & 1.5408(88) & 1.637(10)  \\  
           & 0.1410 & 1.481(12) & 1.582(17) & 1.5341(95) & 1.639(12)  \\  
           & 0.1420 & 1.163(17) & 1.241(21) & 1.515(16)  & 1.617(16)  \\  
\hline   
5.25       & 0.1390 & 1.5509(98) & 1.638(14) & 1.5273(65) & 1.6134(93) \\  
           & 0.1410 & 1.111(13)  & 1.212(19) & 1.5221(97) & 1.661(17)  \\  
\end{tabular}
\end{center}  
\end{table}

\begin{table}[htb]
\caption{\RC\ action combination: baryon masses.
The run marked with (*) is on the $16^3\!\times\!32$ lattice.}
\label{tab:baryonRC}
\vspace{2mm}
\begin{center} 
\begin{tabular}{llllll}
 $\beta$ & $K$ & $m_{\rm N}a$ & $m_{\Delta}a$ & $m_{\rm N}/m_{\rm V}$ & 
 $m_\Delta/m_{\rm V}$ \\
\hline
$1.9^*$      & 0.1370 & 2.195(10)  & 2.296(15) & 1.5547(66) & 1.6263(97) \\ 
             & 0.1400 & 1.845(10)  & 1.978(13) & 1.5428(64) & 1.6541(92) \\ 
             & 0.1420 & 1.494(12)  & 1.662(17) & 1.474(11)  & 1.640(17)  \\ 
             & 0.1430 & 1.283(13)  & 1.501(17) & 1.448(15)  & 1.694(19)  \\ 
             & 0.1435 & 1.154(12)  & 1.368(24) & 1.448(19)  & 1.717(28)  \\ 
             & 0.1440 & 0.972(25)  & 1.171(32) & 1.376(29)  & 1.658(33)  \\ 
\hline
1.9          & 0.1370 & 2.2172(91) & 2.358(20) & 1.5735(61) & 1.673(14) \\ 
             & 0.1400 & 1.8573(95) & 2.009(12) & 1.5434(77) & 1.670(11) \\ 
             & 0.1420 & 1.5195(78) & 1.712(11) & 1.4972(76) & 1.687(11) \\ 
             & 0.1430 & 1.274(11)  & 1.486(13) & 1.431(13)  & 1.669(14) \\
             & 0.1435 & 1.173(22)  & 1.406(39) & 1.463(28)  & 1.754(43) \\
\hline   
$2.0_{\rm tree}$ & 0.1420 & 1.9605(86) & 2.0646(90)& 1.5807(67) & 1.6647(60)\\ 
                 & 0.1450 & 1.6293(87) & 1.733(13) & 1.5644(60) & 1.6644(91)\\ 
                 & 0.1480 & 1.197(15)  & 1.382(25) & 1.485(18)  & 1.714(28) \\ 
                 \cline{2-6} 
$2.0_{\rm pMF}$  & 0.1300 & 2.286(10)  & 2.353(12) & 1.5569(48) & 1.6029(61) \\ 
                 & 0.1370 & 1.4918(78) & 1.622(14) & 1.5220(77) & 1.655(10)  \\ 
                 & 0.1388 & 1.150(16)  & 1.302(26) & 1.487(22)  & 1.684(32)  \\ 
                 \cline{2-6} 
$2.0_{\rm MF}$   & 0.1300 & 2.2242(46) & 2.3057(61)& 1.5471(27) & 1.6038(37) \\ 
                 & 0.1340 & 1.8185(53) & 1.929(12) & 1.5465(42) & 1.6405(92) \\ 
                 & 0.1370 & 1.419(10)  & 1.521(15) & 1.5096(95) & 1.618(13)  \\ 
                 & 0.1388 & 1.153(12)  & 1.308(19) & 1.489(15)  & 1.689(20)  \\ 
\end{tabular}
\end{center}  
\end{table}

\section{String Tension}
\label{appendix:string}

\begin{table}[h]
\caption{Results of string tension $\sigma$ in lattice units.
The quoted error of $\sigma$ includes the estimate of the systematic error
described in Sec.\ref{sec:sigma}.
We also show the fitting range $r_{\rm min}$--$r_{\rm max}$. 
The run marked with (*) is on the $16^3\!\times\!32$ lattice.}
\label{tab:fit2rst}
\begin{center}  
\begin{tabular}{lccccll} 
    \makebox[15mm][c] {action}        &
    \makebox[15mm][c] {$\beta$}       &
    \makebox[15mm][c] {K}             &
    \makebox[15mm][c] {$\sigma$}      &
    \makebox[15mm][c] {$r_{\rm min}$} &
    \makebox[15mm][c] {$r_{\rm max}$} \\ 
\hline
\PW\    & 5.0 & 0.1779 & 0.324(38)   & $2\sqrt{2}$--$2\sqrt{3}$ & 5 \\
        &     & 0.1798 & 0.307(27)   & $2\sqrt{2}$--$2\sqrt{3}$ & 5 \\
        &     & 0.1811 & 0.335(11)   & $2\sqrt{2}$--$2\sqrt{3}$ & 5 \\
\hline
\RW\    & 1.9 & 0.1688 & 0.2980(53)  & $2\sqrt{2}$--$2\sqrt{3}$ & 6 \\
        & 2.0 & 0.1583 & 0.2678(60)  & $2\sqrt{2}$--$2\sqrt{3}$ & 6 \\
        &     & 0.1623 & 0.2143(42)  & $2\sqrt{2}$--$2\sqrt{3}$ & 6 \\
        &     & 0.1644 & 0.1864(42)) & $2\sqrt{2}$--$2\sqrt{3}$ & 6 \\
\hline
\PCmf\  & 5.0 & 0.1415 & 0.338(54)  & $\sqrt{6}$--3 & 5 \\
        &     & 0.1441 & 0.317(35)  & $\sqrt{6}$--3 & 5 \\
        &     & 0.1455 & 0.323(37)  & $\sqrt{6}$--3 & 5 \\
        & 5.2 & 0.1390 & 0.2192(90) & $2\sqrt{2}$--$2\sqrt{3}$ & 6 \\
        &     & 0.1410 & 0.1588(50) & $2\sqrt{2}$--$2\sqrt{3}$ & 6 \\
        &     & 0.1420 & 0.1255(39) & $2\sqrt{2}$--$2\sqrt{3}$ & 6 \\
        & 5.25& 0.1390 & 0.1453(59) & $2\sqrt{2}$--$2\sqrt{3}$ & 6 \\
        &     & 0.1410 & 0.0969(34) & $2\sqrt{2}$--$2\sqrt{3}$ & 6 \\
\hline
\RCpmf\ & 1.9 & 0.1370 & 0.3243(87) & $2\sqrt{2}$--3 & 5 \\
        &     & 0.1400 & 0.2750(75) & $2\sqrt{2}$--3 & 5 \\
        &     & 0.1420 & 0.2465(46) & $2\sqrt{2}$--3 & 5 \\
\hline
\RCpmf & $1.9^*$ & 0.1420 & 0.2375(60) & $2\sqrt{2}$--$2\sqrt{3}$ & 8 \\
       &         & 0.1430 & 0.2094(51) & $2\sqrt{2}$--$2\sqrt{3}$ & 8 \\
       &         & 0.1440 & 0.1755(57) & $2\sqrt{2}$--$2\sqrt{3}$ & $3\sqrt{5}$ \\
\hline
 \RCtr\ & 2.0 & 0.1420 & 0.2583(81) & $2\sqrt{2}$--$2\sqrt{3}$ & 6 \\
        &     & 0.1450 & 0.2097(47) & $2\sqrt{2}$--$2\sqrt{3}$ & 6 \\
        &     & 0.1480 & 0.1642(53) & $2\sqrt{2}$--$2\sqrt{3}$ & 6 \\
\hline
 \RCmf\ & 2.0 & 0.1300 & 0.2147(57) & $2\sqrt{2}$--$2\sqrt{3}$ & 6 \\
        &     & 0.1340 & 0.1832(48) & $2\sqrt{2}$--$2\sqrt{3}$ & 6 \\
        &     & 0.1370 & 0.1506(38) & $2\sqrt{2}$--$2\sqrt{3}$ & 6 \\
        &     & 0.1370 & 0.1251(35) & $2\sqrt{2}$--$2\sqrt{3}$ & 6 \\
\end{tabular}
\end{center}
\end{table}

%%%%%%%%%%%%%% figures %%%%%%%%%%%%%%%%

\newpage

\newpage

\end{document}